\documentclass[12pt,preprint]{aastex}
\slugcomment{} 
\shorttitle{SDSS QUASAR LENS SEARCH. VI.}
\shortauthors{OGURI ET AL.}
\begin{document}
\title{The Sloan Digital Sky Survey Quasar Lens Search. VI. Constraints on 
  Dark Energy and the Evolution of Massive Galaxies}  
%
\author{
Masamune Oguri,\altaffilmark{1,2} 
Naohisa Inada,\altaffilmark{3,4} 
Michael A. Strauss,\altaffilmark{5} 
Christopher S. Kochanek,\altaffilmark{6} 
Issha Kayo,\altaffilmark{1,7} 
Min-Su Shin,\altaffilmark{5,8} 
Tomoki Morokuma,\altaffilmark{9} 
Gordon T. Richards,\altaffilmark{10}
Cristian E. Rusu,\altaffilmark{11,12}
Joshua A. Frieman,\altaffilmark{13,14,15}
Masataka Fukugita,\altaffilmark{1,16}
Donald P. Schneider,\altaffilmark{17,18} 
Donald G. York,\altaffilmark{15,19}
Neta A. Bahcall,\altaffilmark{5} and
Richard L. White\altaffilmark{20}
}

\altaffiltext{1}{Institute for the Physics and Mathematics of the
  Universe, University of Tokyo, 5-1-5 Kashiwanoha, Kashiwa, Chiba
  277-8583, Japan.} 
\altaffiltext{2}{Division of Theoretical Astronomy, National Astronomical
Observatory of Japan, 2-21-1 Osawa, Mitaka, Tokyo 181-8588, Japan.} 
\altaffiltext{3}{Department of Physics, Nara National College of
  Technology, Yamatokohriyama, Nara 639-1080, Japan.}    
\altaffiltext{4}{Research Center for the Early Universe, School of
  Science, University of Tokyo, Bunkyo-ku, Tokyo 113-0033, Japan.}
\altaffiltext{5}{Princeton University Observatory, Peyton Hall,
  Princeton, NJ 08544, USA.}                                   
\altaffiltext{6}{Department of Astronomy, The Ohio State University, 
                  Columbus, OH 43210, USA.}                 
\altaffiltext{7}{Department of Physics, Toho University, Funabashi,
  Chiba 274-8510, Japan.} 
\altaffiltext{8}{Department of Astronomy, University of Michigan, 
  500 Church Street, Ann Arbor, MI 48109-1042 USA.}
\altaffiltext{9}{Institute of Astronomy, School of Science, University
  of Tokyo, 2-21-1 Osawa, Mitaka, Tokyo 181-0015, Japan.}
\altaffiltext{10}{Department of Physics, Drexel University, 3141
  Chestnut Street,  Philadelphia, PA 19104, USA.}
\altaffiltext{11}{Optical and Infrared Astronomy Division, National
  Astronomical Observatory of Japan, 2-21-1 Osawa, Mitaka, Tokyo
  181-8588, Japan.} 
\altaffiltext{12}{Department of Astronomy, Graduate School of Science,
  University of Tokyo 7-3-1, Hongo Bunkyo-ku, Tokyo 113-0033, Japan.}
\altaffiltext{13}{Kavli Institute for Cosmological Physics, University 
  of Chicago, Chicago, IL 60637, USA.}
\altaffiltext{14}{Center for Particle Astrophysics, Fermi National 
  Accelerator Laboratory, P.O. Box 500, Batavia, IL 60510, USA.}
\altaffiltext{15}{Department of Astronomy and Astrophysics, The University 
  of Chicago, 5640 South Ellis Avenue, Chicago, IL 60637, USA.}
\altaffiltext{16}{Institute for Cosmic Ray Research, University of Tokyo, 
  Kashiwa, 277-8582, Japan.} 
\altaffiltext{17}{Department of Astronomy and Astrophysics, The
  Pennsylvania State University, 525 Davey Laboratory, 
  University Park, PA 16802, USA.}   
\altaffiltext{18}{Institute for Gravitation and the Cosmos,The
  Pennsylvania State University, 525 Davey Laboratory, 
  University Park, PA 16802, USA.}   
\altaffiltext{19}{Enrico Fermi Institute, The University of Chicago,
  5640 South Ellis Avenue, Chicago, IL 60637, USA.} 
\altaffiltext{20}{Space Telescope Science Institute, 3700 San Martin
  Drive, Baltimore, MD 21218, USA.} 

\begin{abstract}
We present a statistical analysis of the final lens sample from
the Sloan Digital Sky Survey Quasar Lens Search (SQLS). The number
distribution of a complete subsample of 19 lensed quasars selected
from 50,836 source quasars is compared with theoretical expectations,
with particular attention to the selection function. Assuming that the
velocity function of galaxies does not evolve with redshift, the SQLS
sample constrains the cosmological constant to  
$\Omega_\Lambda=0.79^{+0.06}_{-0.07}({\rm
 stat.})^{+0.06}_{-0.06}({\rm syst.})$ for a flat universe. The
dark energy equation of state is found to be consistent with $w=-1$
when the SQLS is combined with constraints from baryon acoustic
oscillation (BAO) measurements or results from the Wilkinson Microwave
Anisotropy Probe (WMAP). We also obtain simultaneous constraints on
cosmological parameters and redshift evolution of the galaxy
velocity function, finding no evidence for redshift evolution at $z\la
1$ in any combinations of constraints. For instance, number density
evolution quantified as $\nu_n\equiv d\ln\phi_*/d\ln(1+z)$ and the
velocity dispersion evolution $\nu_\sigma\equiv d\ln\sigma_*/d\ln(1+z)$ 
are constrained to  
$\nu_n=1.06^{+1.36}_{-1.39}({\rm stat.}){}^{+0.33}_{-0.64}({\rm syst.})$ and
$\nu_\sigma=-0.05^{+0.19}_{-0.16}({\rm stat.}){}^{+0.03}_{-0.03}({\rm syst.})$ 
respectively when the SQLS result is combined with BAO and WMAP for
flat models with a cosmological constant. We find that a significant
amount of dark energy is preferred even after fully marginalizing over
the galaxy evolution parameters. Thus the statistics of lensed quasars
robustly confirm the accelerated cosmic expansion.
\end{abstract}


\section{Introduction}\label{sec:intro}

The statistics of gravitationally lensed quasars serve as a unique
probe of cosmology \citep{turner84}. An advantage of gravitational
lensing is the simple and robust theoretical basis, which enables
solid predictions of lensing signals for a given mass distribution. 
For instance, the probability that a distant quasar is strongly lensed
by an intervening massive galaxy is known to be sensitive to the
value of the cosmological constant or dark energy 
\citep{turner90,fukugita90}. This result arises because the lensing 
probability is sensitive to the cosmological volume in the
intermediate redshift range of $z\sim 0.2-1$ that is a sensitive
function of dark energy. After many early studies based on small
samples 
\citep[e.g.,][]{fukugita92,maoz93,kochanek96,falco98,chiba99},
recent results are broadly consistent with the standard
$\Lambda$-dominated flat cosmological model
\citep[e.g.,][]{chae02,mitchell05,chae07,oguri08a}. 
Unfortunately, small number statistics remain a significant limitation
for the cosmological results. 

In addition to the number of lenses, the redshift distribution of lens
galaxies contains complementary information on cosmological parameters
\citep{kochanek92}. The differential probability distribution of 
lens redshifts is fairly insensitive to both the source quasar
population and magnification bias. On the other hand, the lens
redshift distribution does depend sensitively on any redshift
evolution in the velocity function of galaxies, and therefore it has
been used to constrain both cosmological parameters and galaxy
evolution 
\citep[e.g.,][]{ofek03,chae03b,matsumoto08,chae10,cao12}.
A caveat is that sample incompleteness, including biased sampling in
the image separation and redshift spaces, can bias the results
significantly \citep{capelo07}.

Statistical samples of strongly lensed quasars have been constructed
at both radio and optical wavelengths. The Cosmic Lens All-Sky Survey
\citep[CLASS;][]{myers03,browne03} represents the largest survey of
strongly lensed quasars in the radio. CLASS identified 22 lens systems
from a sample of $\sim 16500$ flat-spectrum radio sources, of which 13
lenses from $\sim 9000$ radio sources constitute a statistically
well-defined sample. While the effects of dust extinction and
foreground lens galaxies on the selection of strong lens systems are
negligible in the radio, the poorly characterized redshift
distribution of the source population is a major problem
\citep[e.g.,][]{munoz03,mckean07}. Any uncertainties in the mean
redshift of the source population directly translate into a 
systematic error on any dark energy constraint derived from the
statistics of the CLASS lenses.  

Our lens survey, the Sloan Digital Sky Survey Quasar Lens
Search \citep[SQLS;][]{oguri06,oguri08a,inada08a,inada10,inada12}, is
constructed at optical wavelengths and represents the largest survey
for gravitationally lensed quasars to date. The SQLS is built on the
spectroscopic quasar catalog of the Sloan Digital Sky Survey
\citep[SDSS;][]{york00}. The redshifts are known for all the source
quasars, allowing robust comparisons of the observed lensing
probability with theoretical predictions to extract cosmological
information. In addition, the spectroscopic data, and the rich color 
information of the 5-band SDSS imaging data, 
make the identification of the lens candidates quite efficient. 
While the relatively poor seeing of the SDSS images prevent us
from discovering quasar lenses with small image separation ($\la 1''$), 
we can make detailed simulations of lenses to characterize the
selection function of our lens survey \citep[][hereafter Paper I]{oguri06}.

In this paper, we present a detailed statistical analysis of the
final lens sample of the SQLS \citep[][hereafter Paper V]{inada12}
based on SDSS Data Release 7 (DR7). The statistical sample consists of
26 quasar lenses 
selected from 50,836 source quasars in the redshift range $0.6<z<2.2$
with Galactic extinction corrected \citep{schlegel98} magnitudes
brighter than $i=19.1$. Paying particular 
attention to selection effects, we compare the number distribution
of the strong lenses with model predictions to obtain cosmological and
astrophysical information. Previous studies of strong lensing
statistics have focused on either constraining cosmological parameters
with a fixed galaxy evolution model, or vice versa. In this paper we 
consider simultaneous constraints on cosmological parameters and
galaxy evolution parameters with several different priors on the 
cosmological parameters from other cosmological probes. 

This paper is organized as follow. After briefly describing the lens
sample used for the statistical analysis in \S\ref{sec:lens}, we
present our theoretical model in detail in \S\ref{sec:theory}.
Section~\ref{sec:cosmo} presents our main results on cosmological
constraints. Constraints on redshift evolution of the velocity
function are given in \S\ref{sec:galevo}. We discuss the results in 
\S\ref{sec:discussions} and we conclude in \S\ref{sec:summary}. 
We denote the present matter density as
$\Omega_M$, the dark energy density as $\Omega_{\rm DE}$, the equation
of state of dark energy as $w$ (which is assumed to be a constant
throughout the paper), and the normalized Hubble constant as $h =
H_0/(100{\rm km\,s^{-1}\,Mpc^{-1}})$. 
For the special case where dark energy is assumed to be a cosmological
constant ($w=-1$), the dark energy density is denoted as
$\Omega_\Lambda$. 

\section{Lensed Quasar Sample}\label{sec:lens}

The SDSS \citep[SDSS-I and SDSS-II Legacy Survey;][]{york00} is an
imaging and spectroscopic survey covering 10,000 square degrees of
the sky, using a dedicated wide-field 2.5-meter telescope
\citep{gunn06} at the Apache Point Observatory in New Mexico, USA.  
Images taken in five broad-band filters
\citep[$ugriz$,][]{fukugita96,gunn98,doi10} are reduced with an
automated pipeline, leading to an astrometric accuracy better than
about $0\farcs1$ \citep{pier03}, and a photometric zeropoint accuracy
of about 0.01 magnitude over the entire survey area  
\citep{hogg01,smith02,ivezic04,tucker06,padmanabhan08}. 
In addition to imaging, SDSS conducts spectroscopic observations
with a multi-fiber spectrograph covering  3800{\,\AA} to 9200{\,\AA},
with a resolution of $R\sim1800$ for targets selected by the imaging
data \citep{blanton03}. All the data are now publicly available 
\citep{stoughton02,abazajian03,abazajian04,abazajian05,abazajian09,
adelman06,adelman07,adelman08}. 

The final statistical lens sample of the SQLS is based on the SDSS
DR7 quasar catalog containing 105,783
spectroscopically confirmed quasars \citep{schneider10}.
While the quasars are selected using several different selection
criteria, a subsample of quasars with $0.6<z<2.2$ and
Galactic extinction corrected point spread function (PSF) magnitudes
of $i<19.1$ is nearly complete, with the completeness independent of
the SDSS image morphology \citep{richards02,richards06,vandenberk05}.
The SQLS constructs a statistical lens sample based on this subsample
of 50,836 quasars. We identify strong lens candidates from SDSS
images of these sources using two selection algorithms. One examines 
the image morphology to select smaller separation lenses (although
still with $\theta>1''$), and the other identifies companions of
similar colors to select lens candidates at wider image separations
out to $\theta<20''$. We also require that the magnitude difference
between the lensed images is smaller than 1.25~mag, because lenses
with larger magnitude differences are difficult to find in the SDSS
images. Our simulations of SDSS images have shown that our selection
algorithms are nearly complete over the chosen image separation 
($1''<\theta<20''$) and magnitude difference ($|\Delta m| < 1.25$)
ranges (Paper I). The lens candidates are then observed with various
facilities to determine if they are real lens systems or not. The final
lens sample consists of 26 strongly-lensed quasars satisfying the
criteria described above. Paper V supplies more details of the
definition of the source quasar sample, the selection of lens
candidates, and the subsequent observations. 

The 26 quasar lenses in the final SQLS statistical lens sample are
summarized in Table~\ref{tab:sample}. Most of the lens redshifts are
measured spectroscopically, either from direct measurement of the
spectrum of the lens galaxy or using strong absorption lines in the
quasar spectrum in which the redshift of the absorber matches the
inferred lens redshift from the color and the apparent magnitude of
the lens. For lenses without spectroscopic redshifts, we adopt the
lens redshifts inferred from these colors and magnitudes and include
their errors as described below.  

We estimate the $I$-band magnitude of the quasar components,
$I_{\rm QSO}$, from the follow-up high angular-resolution imaging. We
define $I_{\rm QSO}$ in analogy to the PSF magnitude in the SDSS
data, but without the contribution of the flux from the lens
galaxy. We derive $I_{\rm QSO}$ from the magnitude of the brightest
quasar image $I_{\rm bri}$ and the total magnitude $I_{\rm tot}$ in
follow-up imaging. In Paper I  it was shown, using simulations of SDSS
images, that the effective magnification factor of lens systems
relevant for the PSF magnitude is given by 
\begin{equation}
\mu=\bar{\mu}\mu_{\rm tot}+(1-\bar{\mu})\mu_{\rm bri},
\label{eq:mag}
\end{equation}
where
\begin{equation}
\bar{\mu}=\frac{1}{2}\left[1+\tanh\left(1.76-1.78\theta\right)\right],
\end{equation}
the image separation $\theta$ is in arcseconds and $\mu_{\rm tot}$ and
$\mu_{\rm bri}$ are the total magnification and the magnification of
the brightest image, respectively. Given this result, we can estimate
$I_{\rm QSO}$ by 
\begin{equation}
I_{\rm QSO}=-2.5\log\{\bar{\mu}10^{-0.4I_{\rm tot}}+(1-\bar{\mu})
10^{-0.4I_{\rm bri}}\}.
\end{equation}
Both $I_{\rm QSO}$ and the $I$-band magnitude of the lens galaxy,
$I_{\rm gal}$, are summarized in Table~\ref{tab:sample}.
 
As in \citet[][hereafter Paper III]{oguri08a} and \citet[][hereafter
  Paper IV]{inada10}, we consider a subsample 
of the statistical lens sample for our final analysis. First, we
restrict the image separation range to 
$\theta_{\rm min}<\theta<\theta_{\rm max}$ with
$\theta_{\rm min}=1''$ and $\theta_{\rm max}=4''$, since we are
interested in galaxy-scale lensed quasars whose lens potentials are
well approximated by singular isothermal distributions 
\citep[e.g.,][]{rusin05,koopmans06,koopmans09,sonnenfeld12}. 
Second, we require that the lens galaxy is fainter than the quasar
images in the $I$-band, $I_{\rm QSO}<I_{\rm gal}$. This condition is
necessary to ensure that emission from the lens galaxies does not
affect our lens candidate selection and the quasar target selection in 
SDSS. We find that 19 of the 26 lenses pass these additional selection
criteria.   

\section{Theoretical Model}\label{sec:theory}

We mostly follow the methodology described in Paper III to
calculate the lensing probabilities for the SQLS lens sample, although
we include a number of modifications and updates for making more
accurate theoretical predictions. 

\subsection{Lens Potential}

Various observations of galaxy-scale strong lenses have convincingly
shown that the radial mass distribution of lens galaxies is, on
average, well described by the isothermal distribution with
$\rho(r)\propto r^{-2}$
\citep[e.g.,][]{rusin05,koopmans06,koopmans09,sonnenfeld12}.
As in Paper III, we adopt the elliptical extension of the
singular isothermal sphere, the singular isothermal ellipsoid (SIE).
The convergence of the SIE is 
\begin{equation}
\kappa(x,y)=\frac{\theta_{\rm E}\lambda(e)}{2}
\left[\frac{1-e}{(1-e)^2x^2+y^2}\right]^{1/2},
\label{eq:siekappa}
\end{equation}
where $\theta_{\rm E}$ is the Einstein radius and $e$ is the
ellipticity of the mass distribution. The Einstein radius is related
to the galaxy velocity dispersion $\sigma_v$ by  
\begin{equation}
\theta_{\rm E}=4\pi\left(\frac{\sigma_v}{c}\right)^2\frac{D_{\rm
    ls}}{D_{\rm os}},
\label{eq:tein}
\end{equation}
where $D_{\rm ls}$ and $D_{\rm os}$ are the angular diameter distances
from lens to source and from observer to source, respectively. 

The parameter $\lambda(e)$ is the velocity dispersion normalization
factor for non-spherical galaxies. Computing $\lambda(e)$ requires
assumptions about the three-dimensional shapes of lens galaxies
\citep[e.g.,][]{keeton98}.
\citet{chae03} computed $\lambda(e)$ for two extreme cases, galaxies
having either oblate or prolate shapes. In our calculation, we use the
following fitting formulae,
\begin{equation}
\lambda_{\rm obl}(e)\approx \exp\left(0.108\sqrt{e}+0.180e^2+0.797e^5\right),
\end{equation}
for the oblate case, and 
\begin{equation}
\lambda_{\rm pro}(e)\approx 1-0.258e+0.827e^6,
\end{equation}
for the prolate case ($e<e_{\rm max}\approx 0.71158$). In our fiducial
model we assume an equal number of prolate and oblate galaxies to 
compute the dynamical normalization 
\begin{equation}
\lambda(e)=f_{\rm obl}\lambda_{\rm obl}(e)+(1-f_{\rm obl})\lambda_{\rm pro}(e),
\end{equation}
with $f_{\rm obl}=0.5$ as a fiducial value.
We assume that the distribution of the ellipticity is described by 
a Gaussian distribution with peak $\bar{e}$ and standard
deviation $\sigma_e$, but truncated at $e=0$ and $0.9$. Based on the
axis ratio distributions of early-type galaxies in the SDSS
\citep{choi07,pradilla08,bernardi10},  we adopt $\bar{e}=0.25$ and
$\sigma_e=0.2$ as our fiducial parameters, which are slightly
different from the values of $\bar{e}=0.3$ and $ \sigma_e=0.16$
adopted in Paper III. 

In addition to the main lens galaxy, we include a contribution from
line-of-sight density fluctuations to the lens potential in the form
of a constant convergence and shear. While the effect of the external
convergence and shear on the total lensing probability is small, the
effect depends strongly on the image separation, such that the
convergence and shear can have a large impact on lensing probabilities
at larger images separations of $\theta\ga 3''$ \citep{oguri05b,faure09}. In
this paper we employ the probability distributions of convergence and
shear presented by 
\citet{takahashi11}, which have been derived from ray-tracing in
high-resolution $N$-body simulations. A caveat is that the matter
fluctuations toward quasar lens systems may be biased compared to the
fluctuation along the random line-of-sight directions, because the
massive galaxies that are typical of strong lens systems are known to
reside in dense environments \citep[e.g.,][]{treu09,faure11}. Such
correlated matter in the vicinity of the lens 
galaxy can contribute to external convergence and shear  
\citep[e.g.,][]{keeton97,holder03,dalal04,momcheva06,suyu10,fassnacht11}. 
Thus in computing the probability distributions of the convergence
$\kappa_{\rm ext}$ and shear $\gamma_{\rm ext}$ coming from the
line-of-sight matter fluctuations, we assume a source redshift of
$z_s=2$, which is higher than the mean redshift of our lens sample but
is still within its redshift range.

Given the lens potential, we compute the lensing cross section
$\sigma_{\rm lens}$ numerically using the public code 
{\it glafic} \citep{oguri10a}. Specifically we randomly create many
sources for each set of parameters (e.g., the ellipticity, external
convergence and shear), and estimate $\sigma_{\rm lens}$ as  
\begin{equation}
\sigma_{{\rm lens},i}=\int d{\bf u} \frac{\Phi(L/\mu)}{\mu\Phi(L)},
\label{eq:mb}
\end{equation}
over the source plane positions ${\bf u}$ where multiple images are
produced and the flux ratio of faint to bright images is larger than
$10^{-0.5}$ for doubles (the flux ratio cutoff of the statistical lens
sample). The index $i$ indicates the number of multiple images, with
$i=2$ for two-image lenses and $i=4$ for four-image lenses. The
parameter $\mu$ is the magnification factor for each source position,
which is computed using Equation~(\ref{eq:mag}). The magnification
factor and the quasar luminosity function $\Phi(L)$ (see
\S\ref{sec:quasarLF}) are needed to include the effect of the
magnification bias. The cross sections are computed in units of the
Einstein radius $\theta_{\rm E}$ and as a function of dimensionless
image separation defined by $\hat{\theta}=\theta/\theta_{\rm E}$.

\subsection{Velocity Function}\label{sec:vf}

The velocity function of galaxies is an essential part of the
theoretical prediction of the lensing probability. In Paper III,
we adopted the velocity function of early-type galaxies obtained from
the SDSS DR5 data \citep{choi07}. In this paper, we use the velocity
function for galaxies of all types \citep{bernardi10}, rather than
that of early-type galaxies only. One of the reasons for using the
all-type velocity functions for the analysis lies in the difficulty in
making robust morphological classifications of the lens galaxy
population. We note, however, that our sample of strong lenses is 
dominated by early-type galaxies because of the image separation cut
of $\theta>1''$. 

Our fiducial velocity function is derived from galaxies of all types
with the velocity dispersion $\sigma_v>125\,{\rm km\,s^{-1}}$ in the
SDSS DR6 data \citep{bernardi10}. The functional form is 
\begin{equation}
\frac{dn}{d\sigma_v}=\phi_*\left(\frac{\sigma_v}{\sigma_*}\right)^\alpha
\exp\left[
  -\left(\frac{\sigma_v}{\sigma_*}\right)^\beta\right]\frac{\beta}{\Gamma 
(\alpha/\beta)}\frac{d\sigma_v}{\sigma_v},
\label{eq:vf}
\end{equation}
with $\phi=2.611\times 10^{-2}(h/0.7)^3\,{\rm Mpc^{-3}}$,
$\sigma_*=159.6\,{\rm km\,s^{-1}}$, $\alpha=0.41$, and
$\beta=2.59$. For comparison, we also consider the velocity function
of galaxies of all types presented by \citet{chae10}, which is based on
the velocity function measurements for the SDSS DR5 galaxies by
\citet{choi07},
\begin{equation}
\frac{dn}{d\sigma_v}=\phi_*
\left[(1-\epsilon)\left(\frac{\sigma_v}{\sigma_*}\right)^\alpha+\epsilon
\frac{\Gamma(\alpha/\beta)}{\Gamma(\alpha'/\beta)}
\left(\frac{\sigma_v}{\sigma_*}\right)^{\alpha'}\right]
\exp\left[
  -\left(\frac{\sigma_v}{\sigma_*}\right)^\beta\right]\frac{\beta}{\Gamma 
(\alpha/\beta)}\frac{d\sigma_v}{\sigma_v},
\end{equation}
with $\phi=7.4\times 10^{-2}h^3{\rm Mpc^{-3}}$,
$\sigma_*=100.0{\rm km\,s^{-1}}$, $\alpha=0.69$,
$\beta=2.01$, $\alpha'=6.61$, and $\epsilon=0.044$.

These velocity functions were derived from the analysis of
low-redshift $z\sim 0.1$ galaxies in the SDSS. We consider
the possibility of redshift evolution of the velocity function
by allowing $\phi_*$ and $\sigma_*$ to evolve as power-laws with
redshift, 
\begin{equation}
\phi_*\rightarrow \phi_*(1+z)^{\nu_n},
\label{eq:evo_nun}
\end{equation}
\begin{equation}
\sigma_*\rightarrow \sigma_*(1+z)^{\nu_\sigma},
\label{eq:evo_nus}
\end{equation}
where we approximated that values of $\phi_*$ and $\sigma_*$ for the
two velocity functions given above are for $z=0$.
The case with $\nu_n=0$ and $\nu_\sigma=0$ corresponds to the no
evolution model that has been adopted in most of previous analyses of
quasar lens statistics (e.g., Paper III). In what follows, we leave
$\nu_n$ and $\nu_\sigma$ as free parameters to be constrained by the
data, except in \S\ref{sec:cosmo} where we assume no redshift evolution. 

\subsection{Quasar Luminosity Function}\label{sec:quasarLF}

The quasar luminosity function (QLF) is needed to compute the
magnification bias in Equation~(\ref{eq:mb}). We adopt the latest
quasar luminosity function from the combined analysis of the SDSS and
2dF data (2SLAQ) presented  by \citet{croom09}. Specifically we use
the pure luminosity evolution model derived in the redshift range
$0.4<z<2.3$ of 
\begin{equation}
\Phi(M_g,z)=\frac{\Phi_*}{10^{0.4(1-\beta_{\rm h})(M_g-M_g^*)}+
10^{0.4(1-\beta_{\rm l})(M_g-M_g^*)}},
\label{eq:qlf}
\end{equation}
\begin{equation}
M_g^*(z)=M_g^*(0)-2.5(k_1z+k_2z^2),
\end{equation}
with parameters given by ($\beta_{\rm h}$, $\beta_{\rm l}$, $\Phi_*$, 
$M_g^*(0)$, $k_1$, $k_2$)$=$(3.33, 1.42, $1.45\times10^{-6}(h/0.7)^3{\rm
 Mpc}^{-3}{\rm mag}^{-1}$, $-22.18+5\log(h/0.7)$, 1.44, $-0.315$). 
The luminosity function is given in terms of rest-frame $g$-band
absolute magnitudes at $z=2$ (i.e., $M_g=M_g(z=2)$). We convert the QLF to
observed $i$-band apparent magnitudes using the K-correction derived in
\citet{richards06}. Given the fact that the quasar luminosity 
function was derived assuming $\Omega_M=0.3$ and $\Omega_\Lambda=0.7$,
we adopt these cosmological parameters for computing the absolute
magnitudes used to compute the magnification bias in equation
(\ref{eq:mb}) no matter what values of $\Omega_{\rm M}$, $\Omega_{\rm
  DE}$, and $w$ we consider for the remainder of the analysis.  

\subsection{Number of Lensed Quasars}

With the lensing cross section computed by
Equation~(\ref{eq:mb}),  we compute the differential probability that
a source at $z=z_s$ and with the SDSS $i$-band PSF magnitude $i=i_{\rm
  QSO}$ is strongly lensed by a lens galaxy at $z=z_l$ with image
separation $\theta$ as   
\begin{equation}
\frac{d^2p_i}{d\theta dz_l}(z_s,i_{\rm QSO})=C_i(\theta)
 \frac{d^2V}{dz_ld\Omega}\int
 \frac{d\hat{\theta}}{\hat{\theta}}\frac{d\sigma_v}{d\theta_{\rm E}}
 \frac{dn}{d\sigma_v}\theta_{\rm E}^2\frac{d\sigma_{{\rm
 lens},i}}{d\hat{\theta}}S(i_{\rm QSO}), 
\label{eq:dpdtdz}
\end{equation}
where $\theta_{\rm E}$ is the Einstein radius defined in
Equation~(\ref{eq:tein}), $\hat{\theta}=\theta/\theta_{\rm E}$,   
$C_i(\theta)$ is the completeness of our lens candidate
selection estimated from simulations of the SDSS images
(see Paper I, note that $C_i(\theta)\approx1 $ for the range of
$\theta$ of our sample), 
\begin{equation}
\frac{d^2V}{dz_ld\Omega}=\frac{c\,dt}{dz_l}(1+z_l)^3D_{\rm ol}^2,
\end{equation}
and
\begin{equation}
S(i_{\rm QSO})={\rm erfc}\left[\frac{i_{\rm QSO}-\bar{i}_{\rm gal}
(\sigma_v,z_l)}{\sqrt{2}\sigma_i}\right].
\end{equation}
The factor $S(i_{\rm QSO})$ is inserted to take into account the
condition that the lens galaxy must be fainter than the PSF
magnitude of the quasar. The mean galaxy $i$-band magnitude
$\bar{i}_{\rm gal}$ is computed from the velocity dispersion
$\sigma_v$ given the observed correlation (the Faber-Jackson relation)
by \citet{bernardi03}, which includes the luminosity evolution of
galaxies with redshift, together with K-correction from
the \citet{coleman80} template spectrum of an elliptical galaxy. While
we consider all types of galaxies as lenses, we use this relation
because early-type galaxies dominate the population of lens galaxies
in our sample, especially because of the removal of small-separation
lenses $\theta<1''$. A Heaviside step function was adopted for
$S(i_{\rm QSO})$ in Paper III, but here we add a Gaussian scatter. 
We use the typical observed scatter in the Faber-Jackson relation of
$\sigma_i=0.5$.  

Given the probability distribution in Equation~(\ref{eq:dpdtdz}), we
can easily compute probability distributions as a function of the
image separation or the lens redshift as
\begin{equation}
\frac{dp_i}{d\theta}(z_s,i_{\rm QSO})=\int_0^{z_s} dz_l
\frac{d^2p_i}{d\theta dz_l},
\label{eq:dpdt}
\end{equation}
\begin{equation}
\frac{dp_i}{dz_l}(z_s,i_{\rm QSO})=\int_{\theta_{\rm
    min}}^{\theta_{\rm max}} d
\theta\frac{d^2p_i}{d\theta dz_l},
\label{eq:dpdz}
\end{equation}
and the total lensing probability is
\begin{equation}
p_i(z_s,i_{\rm QSO})=\int_{\theta_{\rm min}}^{\theta_{\rm max}}
d\theta\int_0^{z_s} dz_l \frac{d^2p_i}{d\theta dz_l}.
\label{eq:ptot}
\end{equation}
The predicted total number of lensed quasars in our quasar sample is
calculated by counting the number of quasars, weighted by the lensing
probability. To speed up the computation, we follow the procedure
introduced in Paper III to calculate the number of lensed quasars for
each redshift-magnitude bin and then sum over the bins. 
We define the number of source quasars in the redshift range 
$z_{s,j}-\Delta z_s/2<z_s<z_{s,j}+\Delta z_s/2$  and a magnitude range 
$i_{{\rm qso},k}-\Delta i/2<i_{\rm qso}< i_{{\rm qso},k}+\Delta i/2$
to be $N_{\rm qso}(z_{s,j}, i_{{\rm qso},k})$. The number
distributions and the total number of lensed quasars become
\begin{equation}
\frac{dN_i}{d\theta}=
 \sum_{z_{s,j}}\sum_{i_{{\rm qso},k}}N_{\rm qso}( z_{s,j}, i_{{\rm
    qso},k}) \frac{dp_i}{d\theta}(z_{s,j}, i_{{\rm qso},k}),
\label{eq:dndt}
\end{equation}
\begin{equation}
\frac{dN_i}{dz_l}=
 \sum_{z_{s,j}}\sum_{i_{{\rm qso},k}}N_{\rm qso}( z_{s,j}, i_{{\rm
    qso},k}) \frac{dp_i}{dz_l}(z_{s,j}, i_{{\rm qso},k}),
\label{eq:dndzl}
\end{equation}
and
\begin{equation}
N_i=\sum_{z_{s,j}}\sum_{i_{{\rm qso},k}}N_{\rm qso}( z_{s,j}, i_{{\rm
    qso},k}) p_i(z_{s,j}, i_{{\rm qso},k}).
\label{eq:numlens}
\end{equation}
Again, the index $i$ indicates the number of multiple images.
The adopted bin widths of $\Delta z_s=0.1$ and $\Delta i=0.2$ are same
as those in Paper III.

\subsection{Likelihood}

We use the likelihood function introduced by \citet{kochanek93} to 
constrain model parameters,
\begin{equation}
  \ln\mathcal{L} = \sum_{\rm lens} \ln\left(
\frac{d^2p_i}{d\theta dz_l}\right)-(N_2+N_4),
\label{eq:like}
\end{equation}
where $d^2p_i/d\theta dz_l$ is calculated from
Equation~(\ref{eq:dpdtdz}) and $N_2$ (doubles) and $N_4$ (quadruples)
are from Equation~(\ref{eq:numlens}). An important improvement from
Paper III is that we now include the lens redshift distribution as a
constraint, particularly because lens redshifts are successfully
measured for most of the lenses used for the analysis. However, there
are three lens systems whose lens redshifts are still not determined
well (see Table~\ref{tab:sample}). For these lenses we include the
lens redshift uncertainties assuming a Gaussian distribution,
\begin{equation}
\frac{d^2p_i}{d\theta dz_l}(z_s,i_{\rm QSO})\rightarrow
\int dz_l \frac{1}{\sqrt{2\pi}\sigma_z}
\exp\left[-\frac{(z_l-\bar{z})^2}{2\sigma_z^2}\right]
\frac{d^2p_i}{d\theta dz_l}(z_s,i_{\rm QSO}),
\end{equation}
where $\bar{z}$ and $\sigma_z$ are listed in Table~\ref{tab:sample}.
We use the estimator, 
\begin{equation}
  \Delta\chi^2=-2\ln\left(\mathcal{L}/\mathcal{L_{\rm max}}\right),
\end{equation}
to derive best-fit model parameters and confidence limits.

\section{Constraints on Cosmological Parameters}\label{sec:cosmo}

In this section, we constrain cosmological parameters by comparing the
observed lensed quasars in the SQLS DR7 statistical sample with
theoretical model predictions. For now we assume that the velocity
function of galaxies does not evolve with redshift (i.e., $\nu_n=0$ and
$\nu_\sigma=0$ in Equations~(\ref{eq:evo_nun}) and (\ref{eq:evo_nus})), 
although  redshift evolution is considered when we estimate
systematic errors. When necessary, the constraints are combined with
those from baryon acoustic oscillation and cosmic microwave background
anisotropy measurements. 

\subsection{Flat Models with a Cosmological Constant}\label{sec:cosmoflat}

\begin{figure}
\epsscale{.7}
\plotone{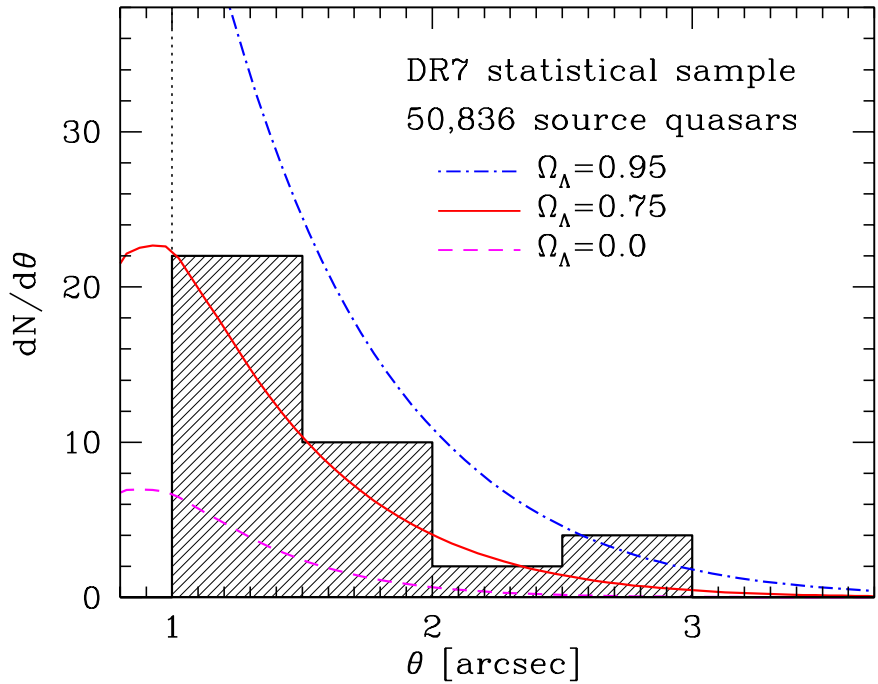}
\caption{The histogram shows the image separation distribution of the
  strong lenses in the statistical lens sample used for our
  cosmological analyses. The subsample contains 19 lenses selected out
  of 50,836 source quasars, as summarized in Table~\ref{tab:sample}.  
  Lines show the theoretical predictions for three different 
  values of the cosmological constant $\Omega_\Lambda$ assuming 
  a flat universe and no evolution of the galaxy velocity
  function. The vertical dotted line shows the $\theta_{\rm
    min}=1''$ lower limit for the image separations in the statistical
  lens sample.  
\label{fig:sep_dr7stat}} 
\end{figure}

\begin{figure}
\epsscale{.7}
\plotone{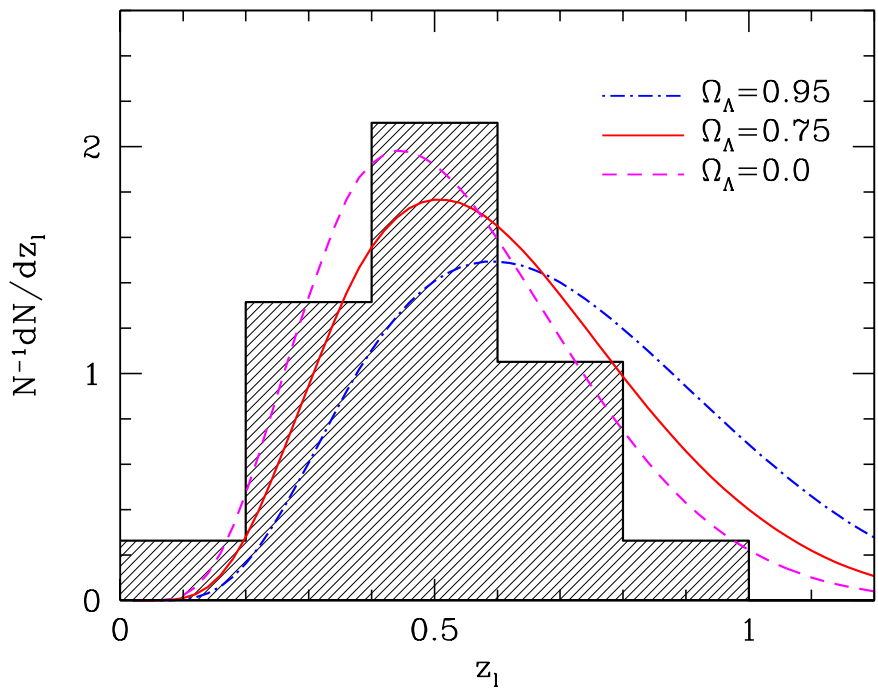}
\caption{The histogram shows the normalized lens redshift distribution
  for our lens sample. Lines show theoretical predictions for
  three different cosmological models, as in
  Figure~\ref{fig:sep_dr7stat}. For lenses with errors on the lens
  redshift (see Table~\ref{tab:sample}), we adopt the best estimated 
  values in constructing the histogram.
  \label{fig:zldist_dr7stat}} 
\end{figure}

We start with the simplest model, a flat universe where dark energy is
described as a cosmological constant $\Omega_\Lambda$ (i.e., $w=-1$). 
This model has only one free parameter, $\Omega_\Lambda$. Before
deriving constraints on $\Omega_\Lambda$, we compare the number
distribution of lenses in our sample as a function of image
separation with the model predictions, as shown in
Figure~\ref{fig:sep_dr7stat}. The total number of lenses is indeed
sensitive to $\Omega_\Lambda$, and models with $\Omega_\Lambda\sim
0.8$ are broadly consistent with the observed number distribution.
In Figure~\ref{fig:zldist_dr7stat}, we examine the normalized lens
redshift distribution $N^{-1}dN/dz_l$. Here we see that the 
observed lens redshift distribution is more consistent with models
with smaller $\Omega_\Lambda$ because of the relatively small number
of lens galaxies at high lens redshifts. 

\begin{figure}
\epsscale{.7}
\plotone{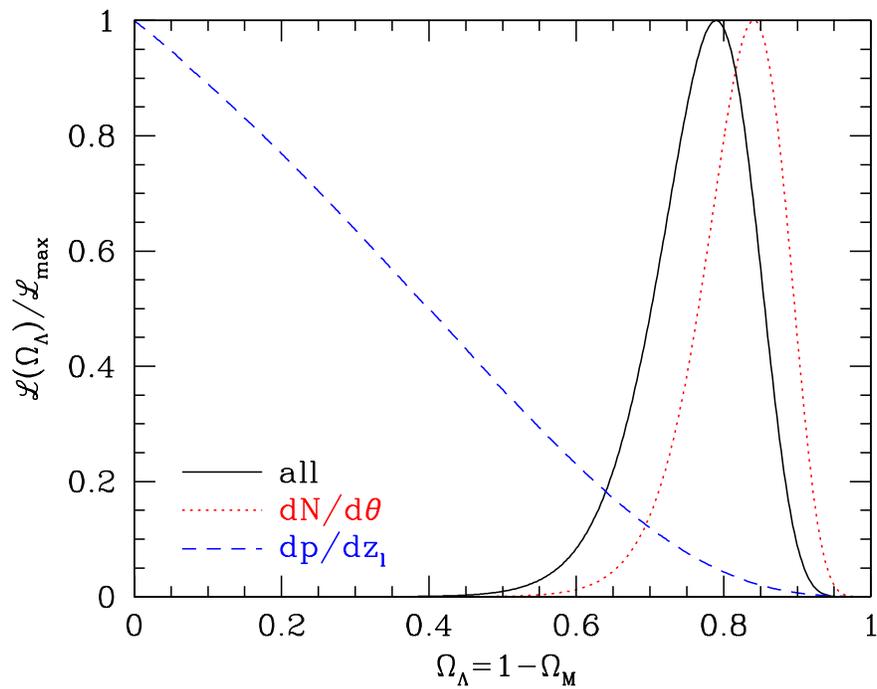}
\caption{Likelihood distributions as a function of the cosmological
  constant $\Omega_\Lambda$ for flat universes. The dotted and dashed
  lines show the separate likelihood distributions for fitting the
  numbers and image separations alone ({\it dotted}) and lens
  redshifts alone ({\it dashed}).
  \label{fig:lambda}} 
\end{figure}

We compute the likelihood of Equation~(\ref{eq:like}) as a function of
$\Omega_\Lambda$. The result shown in Figure~\ref{fig:lambda}
indicates that the SQLS DR7 sample constrains the cosmological
constant to be $\Omega_\Lambda=0.79^{+0.06}_{-0.07}$, where the error
indicates the statistical $1\sigma$ confidence limit. This is consistent
with our previous results presented in Paper III and \citet[][Paper
  IV]{inada10}. The model with  $\Omega_\Lambda=0$ is rejected at
$6\sigma$ level. Dark energy is detected at high significance by
the quasar lens statistics.

If we divide the likelihood into the part contributed by the numbers
and separations as compared to the lens redshifts, we see that there
is some tension. The numbers and separations, which we used in our
previous studies, favor somewhat higher $\Omega_\Lambda$ than the lens
redshifts. In Figure~\ref{fig:lambda}, we show the resulting
likelihoods as a function of $\Omega_\Lambda$ based only on the
observed numbers and separations of lenses, as well as only on the
lens redshift distribution. While the tension is only at
about the $2\sigma$ level, this result may be suggestive of 
redshift evolution in the velocity function. This is one of
the reasons that we consider simultaneous constraints on cosmological
parameters and galaxy evolution in \S\ref{sec:galevo}.

\subsection{Non-Flat Models with a Cosmological Constant}

\begin{figure}
\epsscale{.55}
\plotone{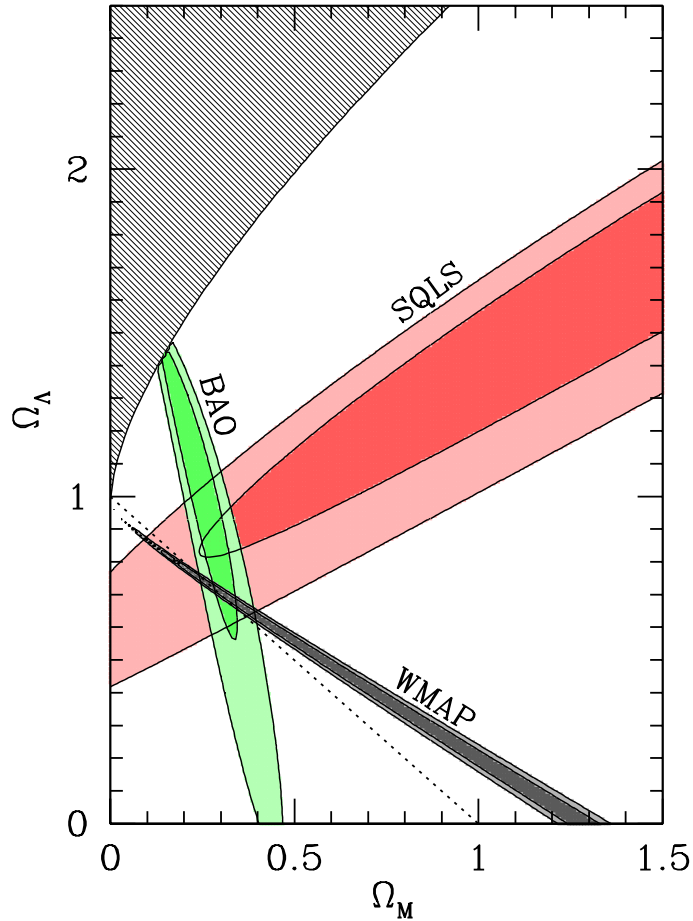}
\caption{Constraints on $\Omega_M$ and $\Omega_\Lambda$ for the
  non-flat models with a cosmological constant. Contours show $1\sigma$ and
  $2\sigma$ confidence regions from the three different cosmological
  probes: SQLS strong lens statistics (this paper), baryon acoustic
  oscillation (BAO) measurements \citep{percival10,blake11}, and the
  CMB anisotropy from WMAP \citep{komatsu11}. The dotted line indicates 
  a flat universe with $\Omega_M+\Omega_\Lambda=1$. The upper left
  shaded region indicates models with no big bang. 
  \label{fig:cont_noflat}} 
\end{figure}

Next we relax the assumption of a flat universe, and consider
cosmological constraints in the $\Omega_M$-$\Omega_\Lambda$ plane.
Figure~\ref{fig:cont_noflat} shows the constraint from the SQLS
DR7 using the full likelihood model. As already known
\citep[e.g.,][]{kochanek96,chae02}, the degeneracy direction of lens
constraints in the $\Omega_M$-$\Omega_\Lambda$ plane resembles that
from Type~Ia supernovae \citep[see, e.g.,][for a recent
result]{suzuki12}. We find that a cosmological constant is
required at the $4\sigma$ level even for this non-flat case.

Because of different degeneracy directions, cosmological parameters
are better constrained by combining several different cosmological
probes. In this paper, we combine our constraints with either those
from the baryon acoustic oscillation (BAO) measurement or from the
cosmic microwave background (CMB). The former uses the baryon wiggle
in the matter power spectrum as a standard ruler. We adopt results
from the WiggleZ Dark Energy Survey \citep{blake11}, which measures 
the BAO scale at $z=0.6$, combined with BAO measurements in the SDSS
luminous red galaxies at $z=0.2$ and $0.35$ \citep{percival10}. 
We consider the anisotropy measured by the Wilkinson Microwave
Anisotropy Probe (WMAP) as the latter. Specifically we consider the
seven-year WMAP result by \citet{komatsu11}, and compute the
likelihood for each cosmological parameter set by the so-called
``distance prior'' which encapsulates all the key information relevant
for dark energy studies derived from the WMAP data. While our lensing
constraints from the SQLS DR7 do not have any dependence on the Hubble
constant, both the BAO and WMAP constraints are sensitive to the
adopted Hubble constant. Thus, when adding the BAO and WMAP we always
include the Hubble constant as an additional free parameter over which
we marginalize to obtain constraints on parameters of interest.  

These BAO and WMAP constraints are also shown in
Figure~\ref{fig:cont_noflat}. The best-fit parameters and $1\sigma$
statistical errors are $\Omega_M=0.28^{+0.03}_{-0.03}$ and
$\Omega_\Lambda=0.88^{+0.09}_{-0.10}$ when the SQLS
is combined with BAO, and $\Omega_M=0.20^{+0.08}_{-0.06}$ and
$\Omega_\Lambda=0.78^{+0.05}_{-0.06}$ when combined with WMAP.
The three constraints are complementary in the sense that their
degeneracy directions are quite different from one another and the
combined constraints give considerably stronger constraints in the
$\Omega_M$-$\Omega_\Lambda$ plane than any of the individual
constraints. That our lensing constraints are consistent with
both the BAO and WMAP constraints is an important cross check of
the current standard cosmological model.  

\subsection{Flat Dark Energy Models}

\begin{figure}
\epsscale{.6}
\plotone{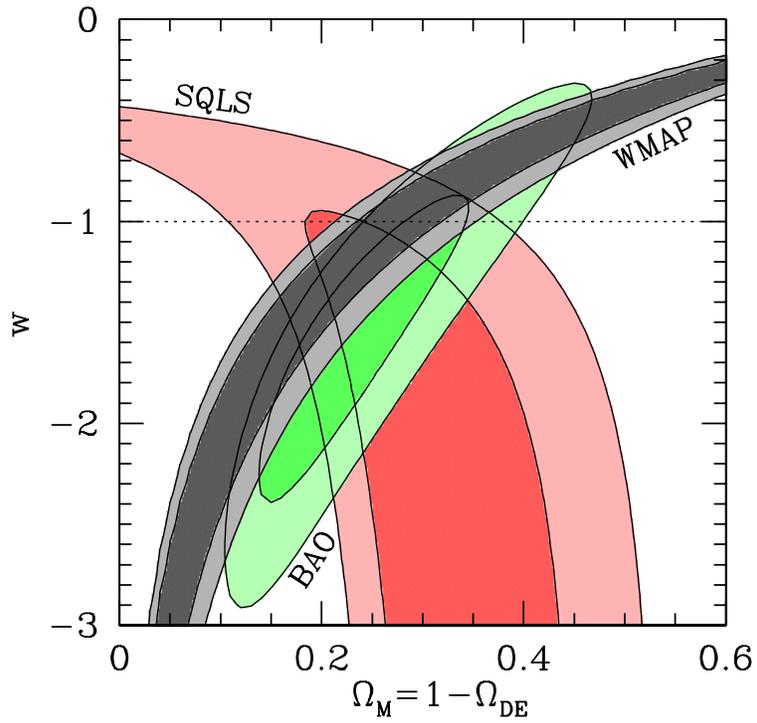}
\caption{Constraints on $\Omega_M=1-\Omega_{\rm DE}$ and
  $w$ for the flat dark energy models. As in 
  Figure~\ref{fig:cont_noflat}, the $1\sigma$ and $2\sigma$
  constraints from the three different cosmological probes are shown
  by contours. The horizontal dotted line indicates a cosmological
  constant with $w=-1$. 
  \label{fig:cont_w}} 
\end{figure}

Next we consider flat models where the dark energy equation of state
$w$ is allowed to vary. This model is parametrized by
$\Omega_M$($=1-\Omega_{\rm DE}$) and $w$. Figure~\ref{fig:cont_w}
shows constraints from the SQLS DR7 as well as those from BAO and
WMAP. The degeneracy direction of our lensing constraint in this plane
is again similar to that of Type~Ia supernovae, and hence is
complementary to the BAO and WMAP constraints. The combination of these
constraints suggests a cosmological model with $\Omega_M\sim 0.3$ and
$w\sim -1$.  

\begin{figure}
\epsscale{.6}
\plotone{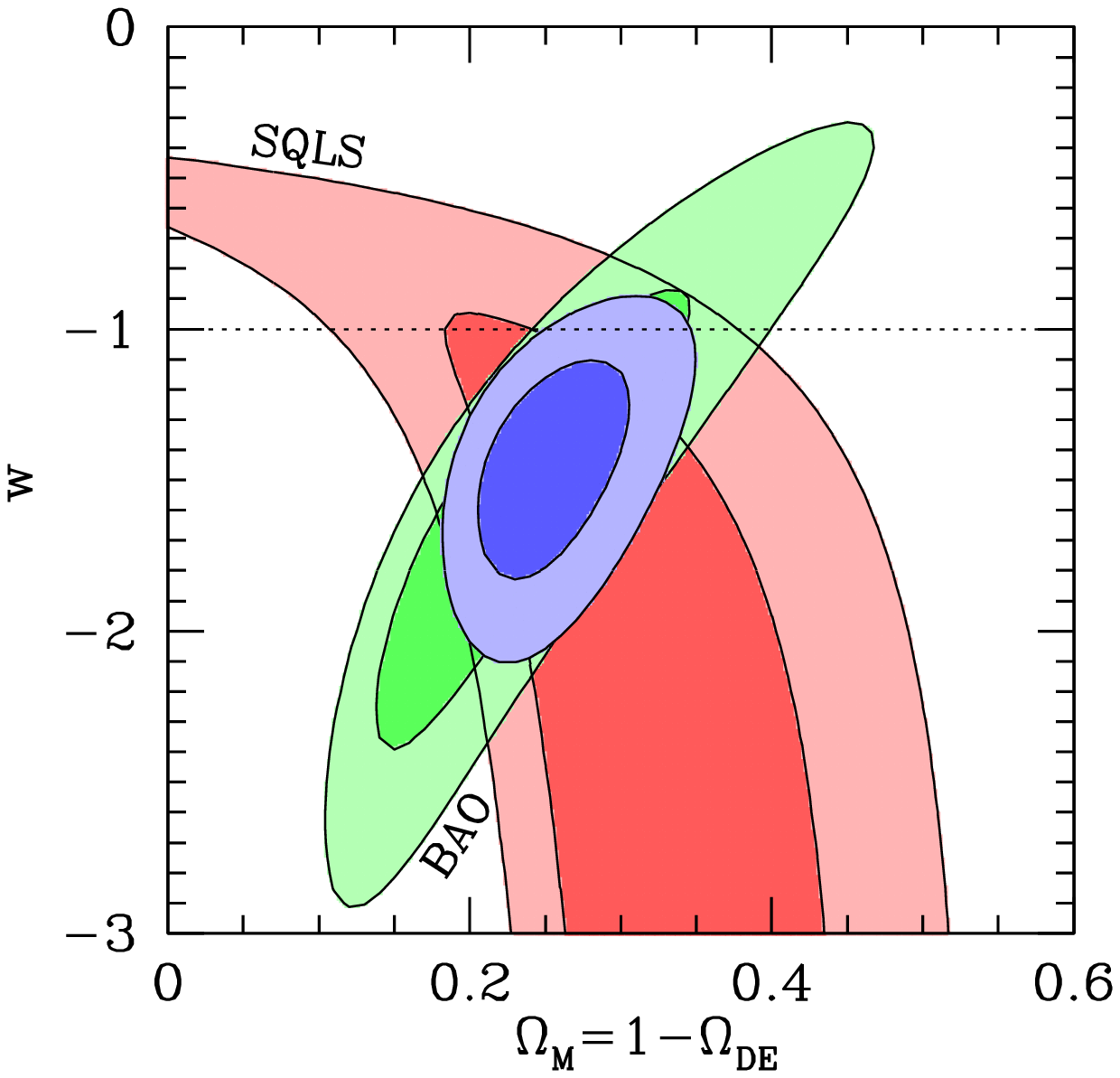}
\caption{Similar to Figure~\ref{fig:cont_w}, but with constraints only
  from the SQLS DR7 and BAO. The innermost contours show the combined
  constraint.  
  \label{fig:cont_w_bao}} 
\end{figure}

\begin{figure}
\epsscale{.6}
\plotone{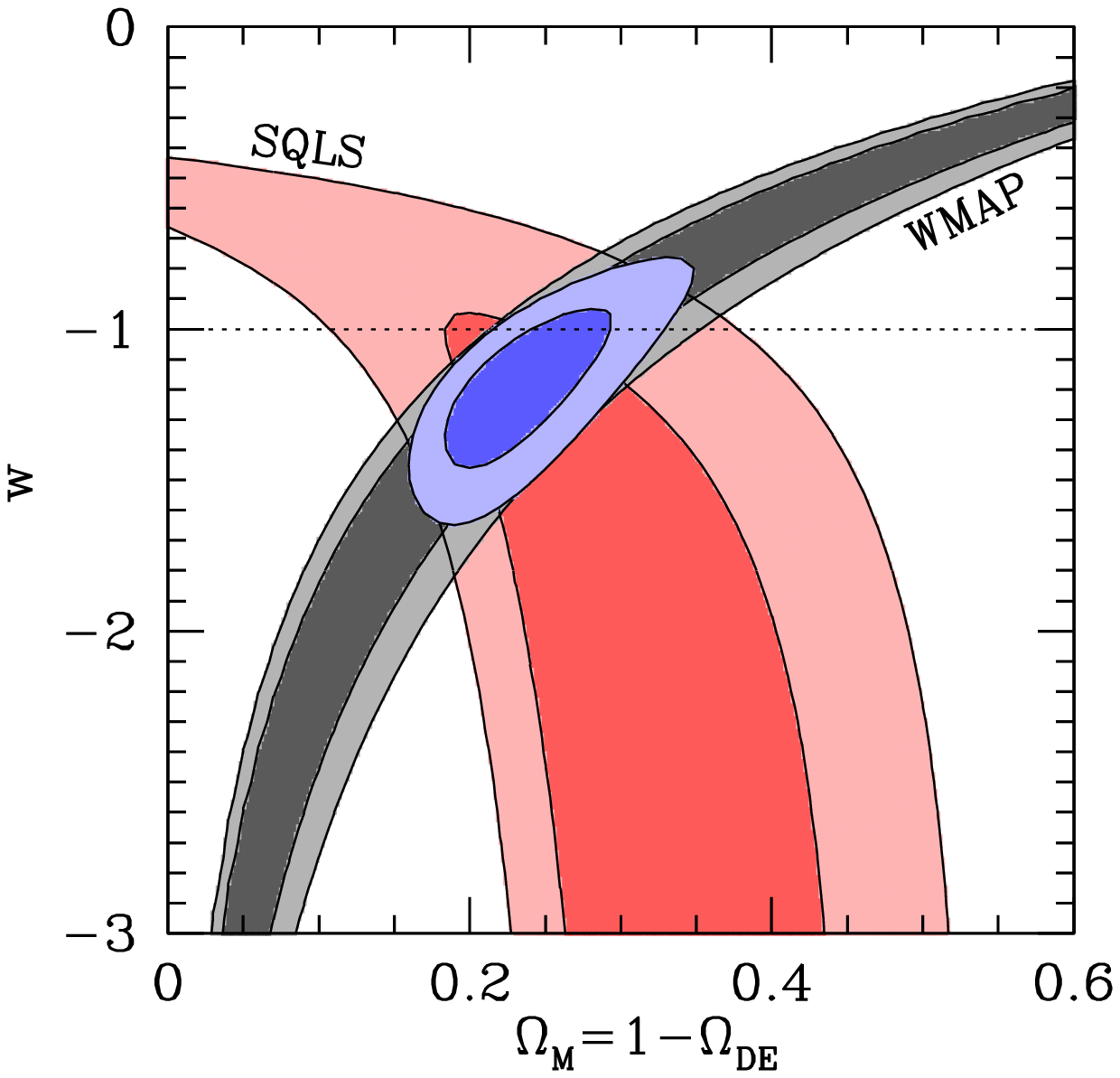}
\caption{Similar to Figure~\ref{fig:cont_w}, but with constraints only
  from the SQLS DR7 and WMAP. The innermost contours show the
  combined constraint. 
  \label{fig:cont_w_wmap}} 
\end{figure}

We also show how $\Omega_M$ and $w$ are constrained when the lensing
information is combined with either the BAO or WMAP result in 
Figure~\ref{fig:cont_w_bao} and Figure~\ref{fig:cont_w_wmap},
respectively. For the former case, the resulting constraints are 
$\Omega_M=0.25^{+0.03}_{-0.03}$ and $w=-1.44^{+0.22}_{-0.25}$, and for
the latter case $\Omega_M=0.23^{+0.04}_{-0.03}$ and 
$w=-1.19^{+0.17}_{-0.17}$. In both cases, $\Omega_M$ and $w$ are
reasonably well constrained thanks to the different degeneracy
directions of the tests.

\subsection{Systematic Errors}\label{sec:syst}

Thus far we have considered only statistical errors. Our model
involves several uncertainties and assumptions which act as systematic
errors in our cosmological analysis. Here we estimate these systematic
errors in a similar way as done in Paper III. Specifically we
consider the following sources of systematic errors.

\begin{itemize}
\item We computed the dynamical normalization $\lambda(e)$ in 
      Equation~(\ref{eq:siekappa}) assuming galaxies consist of
      oblate and prolate populations with equal fractions 
      $f_{\rm obl}=0.5$. As in Paper III, we change the fraction
      by $\pm0.25$ to estimate the uncertainty associated with this
      assumption. 
\item The ellipticity distribution of lens galaxies is assumed to
      be a truncated Gaussian with a peak $\bar{e}=0.25$ and a width
      $\sigma_e=0.2$. We shift the peak $\bar{e}$ by $\pm0.1$ without
      changing the dispersion in order to see how the
      ellipticity distribution affects the cosmological results.
\item We included the line-of-sight convergence $\kappa_{\rm
      ext}$ and shear $\gamma_{\rm ext}$ using
      the PDFs derived from ray-tracing of $N$-body simulations
      \citep{takahashi11} for a fixed source redshift of $z_s=2$.
      We change the assumed redshift by $\pm1$ to estimate the systematic
      error. 
\item The faint end slope of the quasar luminosity function
      in Equation~(\ref{eq:qlf}) is needed for computing the magnification
      bias, yet current measurements of the slope are fairly
      uncertain. Considering results of measurements on the
      quasar luminosity function \citep[e.g.,][]{hopikins07}, we
      change the faint end slope $\beta_{\rm l}$  
      by $\pm0.1$ while fixing the other parameters of the quasar
      luminosity functions. While the range of $\beta_{\rm l}$
      considered here is smaller than what was adopted in
      Paper III, it is still much larger than the measurement
      uncertainty reported in \citet{croom09}.  
\item The velocity function given by Equation~(\ref{eq:vf}) is another
      important 
      source of systematic error. While we adopted the velocity
      function measurement in the SDSS by \citet{bernardi10} as our
      fiducial model, we investigate how the cosmological results are
      altered by adopting the velocity function measurement by
      \citet{chae10}. The specific forms of both the velocity functions
      were given in \S\ref{sec:vf}.
\item While we made the assumption that the velocity function
      does not evolve with redshift, i.e., $\nu_n=0$ and
      $\nu_\sigma=0$ in Equations (\ref{eq:evo_nun}) and
      (\ref{eq:evo_nus}), respectively, we check the effect of 
      redshift evolution by adopting the evolution of
      $\nu_n=-0.229$ and $\nu_\sigma=-0.01$ predicted by the
      semi-analytic model of \citet{kang05}. Note that we will explore
      the effect of letting $\nu_n$ and $\nu_\sigma$ be free
      parameters in \S~\ref{sec:galevo}.
\item As discussed in Paper III, the condition $i_{\rm
      QSO}-i_{\rm gal}<0$  is arbitrary. To estimate the systematic
    error, we shift the  condition to $i_{\rm QSO}-i_{\rm gal}<\pm0.25$, within which 
      the lens sample used for the statistics does not change.
\end{itemize}

In Table~\ref{tab:syst}, we show the contribution of each of these
systematic errors to the systematic error on $\Omega_\Lambda$ for the
flat models with a cosmological constant. We find 
that the largest sources of systemic error are the dynamical
normalization $f_{\rm obl}$ and the velocity function $dn/d\sigma_v$,
followed by the faint end slope of the quasar luminosity function. 
The finding is consistent with the earlier results in Paper III.
The resulting final constraint on $\Omega_\Lambda$ from the SQLS alone
including the systematic error is $\Omega_\Lambda=0.79^{+0.06}_{-0.07}({\rm
  stat.})^{+0.06}_{-0.06}({\rm syst.})$.
Table~\ref{tab:cosmo} summarizes the cosmological constraints for the
three cases studied above, including our estimates of
systematic errors. All the results remain consistent with
the current standard cosmological model ($\Omega_M\sim0.3$,
$\Omega_{\rm DE}\sim 0.7$, and $w\sim -1$).

\section{Evolution of the Velocity Function}\label{sec:galevo}

Thus far we have concentrated on the constraints on cosmological
parameters from the statistics of lensed quasars. However, the lensing
statistics also allow us to study the evolution and structure of
the galaxies that act as lenses. In particular, the lens statistics serve
as a useful probe of the velocity function of massive galaxies at
intermediate redshifts, $0.2\la z\la 1$, which are difficult to
observe directly. Indeed, the analysis in \S\ref{sec:syst} suggests
that redshift evolution of the velocity function is one of the
most significant sources of systematic error. Here we relax the
assumption about the absence of redshift evolution in the velocity
function, and investigate redshift evolution parametrized
by $\nu_n$ (Equation~(\ref{eq:evo_nun})) and $\nu_\sigma$ 
(Equation~(\ref{eq:evo_nus})). Unlike previous studies, however, we still
allow cosmological parameters to vary and consider simultaneous
constraints on the galaxy evolution and cosmological parameters with
the help of external cosmological probes such as BAO and WMAP. 

\begin{figure}
\epsscale{.55}
\plotone{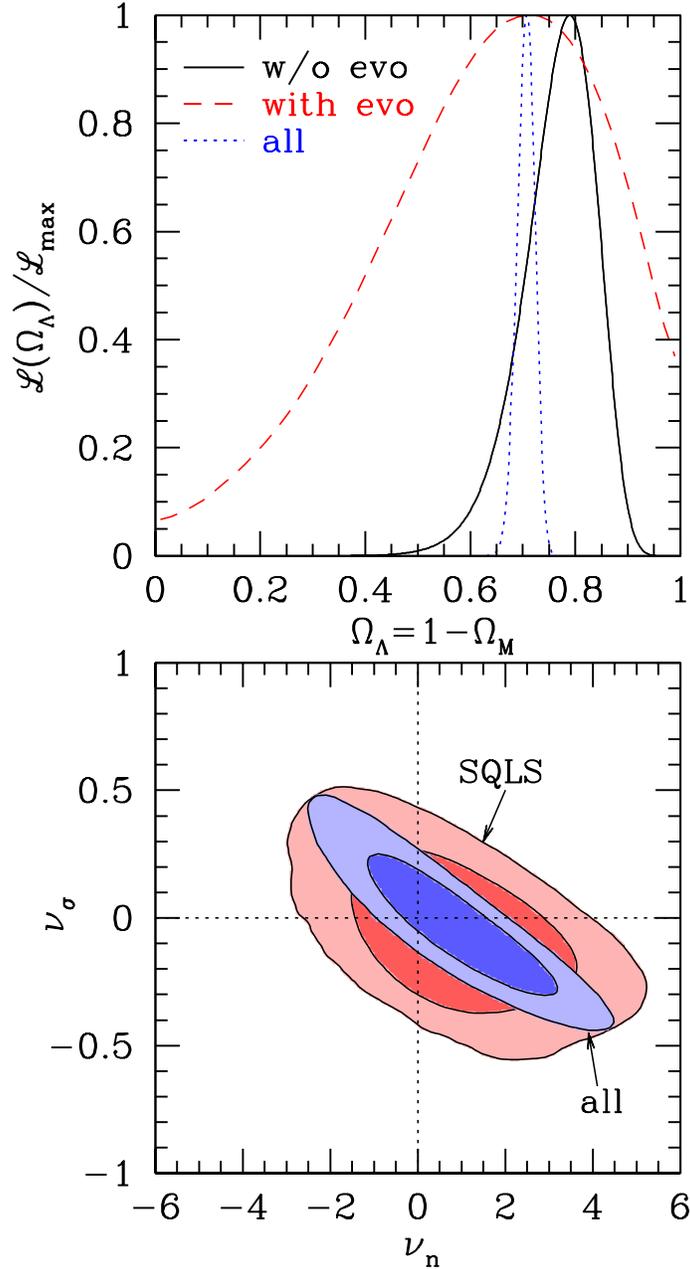}
\caption{Simultaneous constraints on cosmological parameters and 
redshift evolution of the velocity function of galaxies in the flat
models with a cosmological constant. {\it Upper:} Likelihood 
distributions as a function of cosmological constant $\Omega_\Lambda$.
The solid line shows the likelihood from the SQLS alone with no
redshift evolution as in Figure~\ref{fig:lambda}. The dashed line
is the likelihood from the SQLS alone after marginalizing over the
evolution parameters $\nu_n$ and $\nu_\sigma$. The dotted line
is the likelihood distribution marginalizing over $\nu_n$ and $\nu_\sigma$ 
when the SQLS result is combined with the BAO and WMAP constraints. 
{\it Lower:} Constraints on redshift evolution
in the $\nu_n$-$\nu_\sigma$  plane after marginalizing over
$\Omega_\Lambda$. The outer contours are from the SQLS only, and the
inner contours show the combined constraints from SQLS, BAO, and WMAP.
Dotted lines in the lower panel indicate no redshift evolution
($\nu_n=0$ and $\nu_\sigma=0$).  
  \label{fig:cont_evo_flat}} 
\end{figure}

First we consider the flat models with a cosmological constant. With
the two additional parameters $\nu_n$ and $\nu_\sigma$, this model now
has three parameters. The results using SQLS alone are shown in
Figure~\ref{fig:cont_evo_flat}. We find that a cosmological constant
is still required even after the evolution parameters are left as free
parameters and are fully marginalized over. Specifically, the
cosmological constant is constrained to be
$\Omega_\Lambda=0.71^{+0.20}_{-0.25}({\rm stat.}) 
{}^{+0.12}_{-0.10}({\rm syst.})$ from the SQLS
alone, which is consistent with the result assuming no redshift
evolution of the velocity function shown in \S\ref{sec:cosmoflat}. The
model without cosmological constant ($\Omega_\Lambda=0$) is still
inconsistent with the data at more than $2\sigma$. The
constraint projected in the $\nu_n$-$\nu_\sigma$ plane also indicates
that the SQLS data are consistent with the no-redshift-evolution case
($\nu_n=\nu_\sigma=0$) at $1\sigma$. Also note that the most
degenerate direction in the evolution parameters roughly corresponds
to no evolution in the lensing optical depth. The additional
constraints from BAO and WMAP, which significantly improve the
constraint on $\Omega_\Lambda$, similarly improve the constraints in
the $\nu_n$-$\nu_\sigma$ plane. With these additional constraints,
results are still consistent with the no-redshift-evolution
case. Specifically, the measured values of the two parameters are 
$\nu_n=1.06^{+1.36}_{-1.39}({\rm stat.}){}^{+0.33}_{-0.64}({\rm syst.})$ and
$\nu_\sigma=-0.05^{+0.19}_{-0.16}({\rm stat.}){}^{+0.03}_{-0.03}({\rm syst.})$ 
when all three constraints are combined. The systematic errors are
estimated in the same way as in \S\ref{sec:syst}, including the
sources of the systematic error other than redshift evolution of
the velocity function. 

\begin{figure}
\epsscale{.55}
\plotone{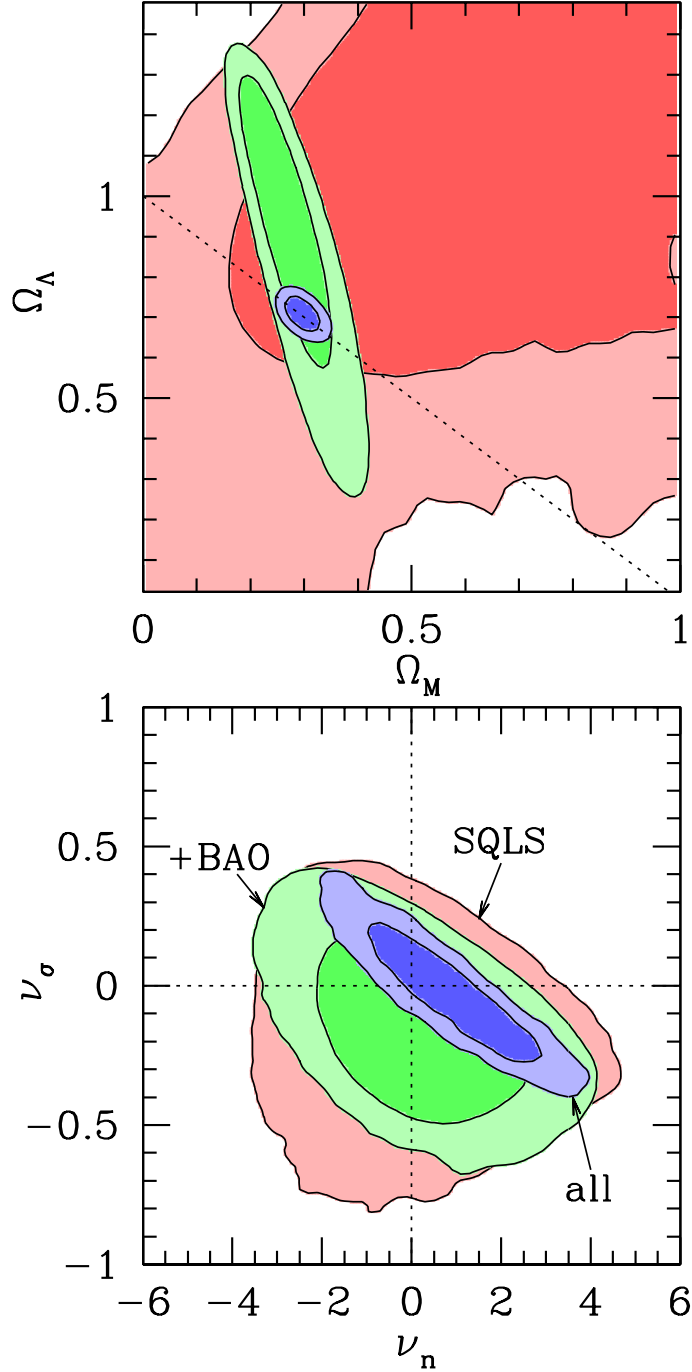}
\caption{Simultaneous constraints on cosmological parameters and 
redshift evolution of the velocity function of galaxies in the
non-flat models with a cosmological constant. The upper panel 
shows constraints in the $\Omega_M$-$\Omega_\Lambda$ plane after
marginalizing over the evolution parameters $\nu_n$ and $\nu_\sigma$.
The dotted line is a flat universe with
$\Omega_M+\Omega_\Lambda=1$. 
Lower panel shows constraints in the $\nu_n$-$\nu_\sigma$ after
marginalizing over the cosmological parameters. In both panels, from
outer to inner contours, we show constraints from the SQLS alone, SQLS
plus BAO, and the combination of all three probes (SQLS+BAO+WMAP),
respectively. Dotted lines in the lower panel indicate no redshift
evolution ($\nu_n=0$ and $\nu_\sigma=0$).  
  \label{fig:cont_evo_noflat}} 
\end{figure}

\begin{figure}
\epsscale{.55}
\plotone{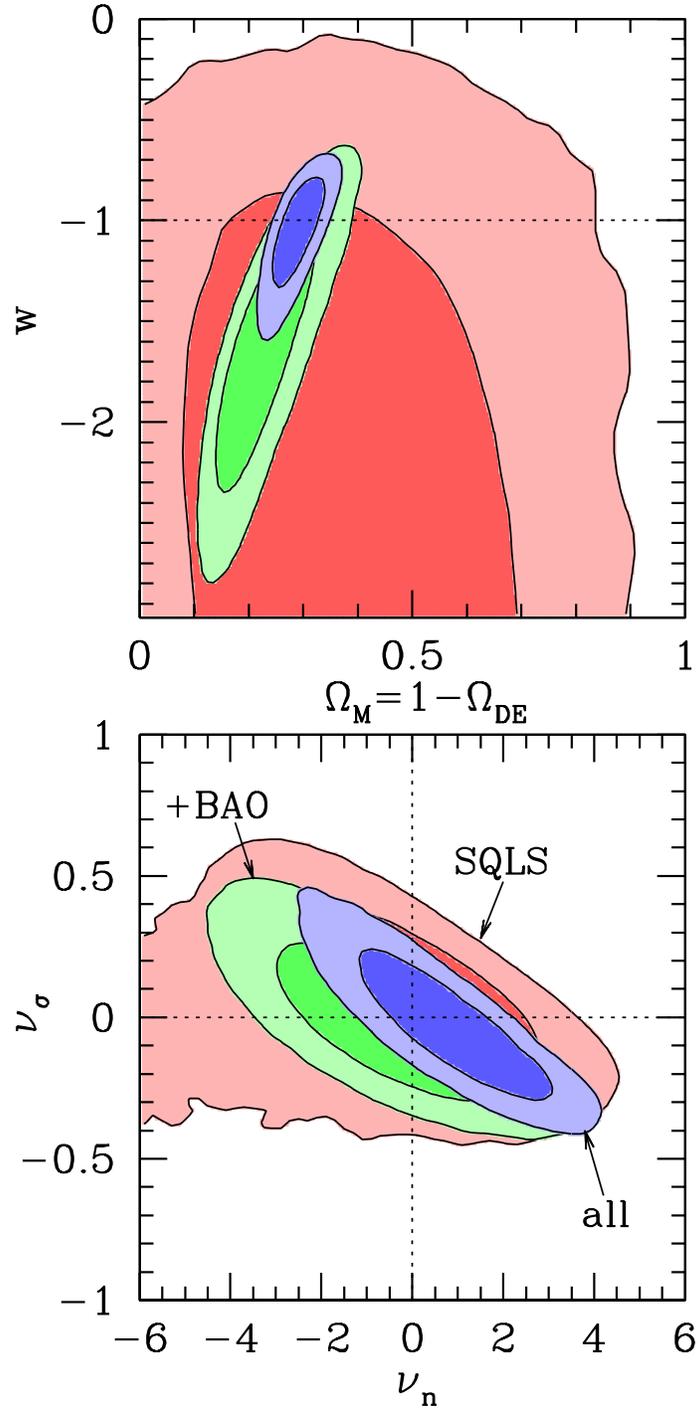}
\caption{As in Figure~\ref{fig:cont_evo_noflat}, but for flat
  dark energy models with $w$ as a free parameter.
  \label{fig:cont_evo_w}} 
\end{figure}

We conduct similar analyses of the simultaneous constraints in the
non-flat models with a cosmological constant and in the flat dark
energy models. The results, shown in Figures~\ref{fig:cont_evo_noflat} and  
\ref{fig:cont_evo_w}, indicate that the cosmological constraints from
SQLS become much weaker when we allow the evolution parameters to vary,
although some useful constraints are still obtained. For instance, in
the non-flat cosmological constant case, non-zero $\Omega_\Lambda$
is again preferred at about $2\sigma$. On the other hand, the
evolution parameters are constrained reasonably well even after
marginalizing over cosmological parameters. In all the cases, the SQLS
data are consistent with no redshift evolution, which supports early
($z\ga 1$) formation and passive evolution of massive early-type
galaxies. We give a summary of the constraints in Table~\ref{tab:evo}.
Note that systematic errors on the cosmological parameters appear
smaller than in Table~\ref{tab:cosmo}, because of weaker cosmological
constraints from the SQLS after marginalizing over the evolution
parameters. 

Our non-detection of redshift evolution of the
velocity function is consistent with earlier results by
\citet{mao94}, \citet{ofek03}, \citet{chae03b}, \citet{capelo07}, and
\citet{matsumoto08}, although we believe our 
result is more robust given the carefully controlled lens sample from
the SQLS and the exploration of the degeneracy with cosmological
parameters. On the other hand, \citet{chae10} reported that the lens
data imply redshift evolution of the velocity function once a more
complicated model is adopted for the evolution. To check this result,
we consider the evolution of the shape of the velocity function by
replacing $\alpha$ and $\beta$ in Equation~(\ref{eq:vf}) as
\citet{chae10} has suggested,
\begin{equation}
\alpha\rightarrow \alpha\left(1+k_\beta\frac{z}{1+z}\right),
\end{equation}
\begin{equation}
\beta\rightarrow \beta\left(1+k_\beta\frac{z}{1+z}\right).
\end{equation}
These additional parameterizations lead to differential
redshift evolution in the number density of galaxies with different
velocity dispersions. In particular, redshift evolution naively
expected from the redshift dependence of the halo mass function, which
predicts stronger  redshift evolution for larger velocity dispersions
\citep{mitchell05,matsumoto08}, can well be described by this
parametrization \citep{chae10}.
 
\begin{figure}
\epsscale{.55}
\plotone{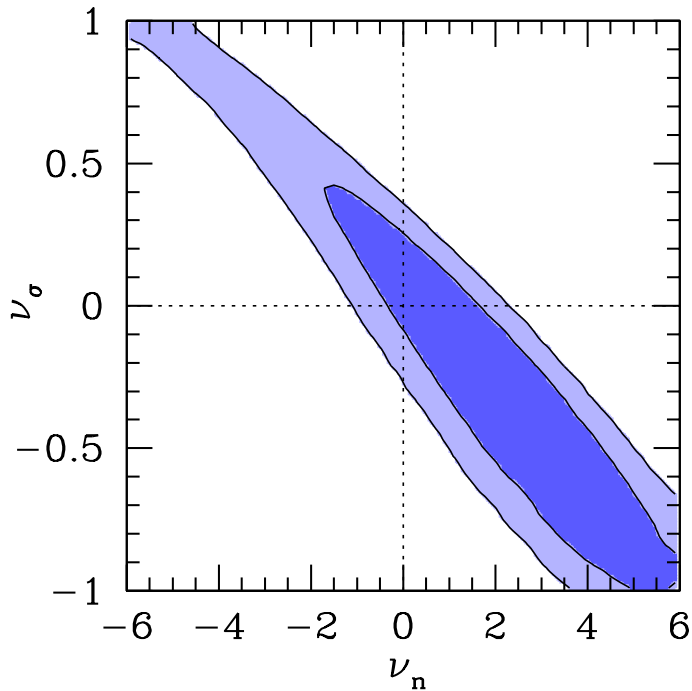}
\caption{Constraints on redshift evolution of the velocity
  function. Here we include the additional parameter $k_\beta$ which 
  describes  redshift evolution of the 
  shape of the velocity function as proposed by \citet{chae10}.
  We consider the flat models with a cosmological constant. The SQLS
  constraint is combined with BAO and WMAP. We show constraints in the
  $\nu_n$-$\nu_\sigma$ after marginalizing over the other parameters
  $\Omega_\Lambda$ and $k_\beta$. Dotted lines in the lower panel
  indicate no redshift evolution ($\nu_n=0$ and $\nu_\sigma=0$).  The
  marginalized constraint on $k_\beta$ is $k_\beta=-0.35^{+0.67}_{-0.35}$.
  \label{fig:cont_evo_flat_ext}} 
\end{figure}

Figure~\ref{fig:cont_evo_flat_ext} shows constraints in the
$\nu_n$-$\nu_\sigma$ plane in the flat models with a cosmological
constant using all three cosmological constraints and marginalizing
over the additional evolution parameter $k_\beta$ as well 
as the cosmological constant $\Omega_\Lambda$. We find no redshift
evolution even in this case. The constraints on the individual 
evolution parameters are $\nu_n=2.47^{+2.27}_{-2.77}$,
$\nu_\sigma=-0.35^{+0.52}_{-0.49}$, and
$k_\beta=-0.35^{+0.67}_{-0.35}$ (statistical errors only), which are
fully consistent with the fiducial no-redshift-evolution case of
$\nu_n=\nu_\sigma=k_\beta=0$. Our different conclusion might be due to
the fact that our source quasar sample is restricted to $z_s<2.2$ and
therefore probes relatively low lens redshifts of $z_l\la1$, whereas
the lens sample used by \cite{chae10} extends to lens redshifts of
$z_l>1$. On the other hand, our SQLS sample is more complete and
has a better understood selection function than the somewhat
heterogeneous, incomplete sample of source and lens redshifts adopted 
by \citet{chae10}.

\section{Discussions}\label{sec:discussions}

\subsection{Comparisons with Other Studies on Galaxy Evolution}

While our results are consistent with no evolution, the constraints
are relatively weak. In the $\nu_n$-$\nu_\sigma$ plane, the evolution
is roughly constrained to keep the lensing optical depth ($\propto
\phi_*\sigma_*^4$) constant while weakly constraining the orthogonal
combination. Studies of galaxy evolution have typically found
significant number and mass evolution in the early-type population
from $z=0$ to $1$ with a decline in the  abundance of early-type
galaxies by roughly a factor of $2$ by $z=1$ \cite[e.g.,][]{faber07,brown07}, 
which corresponds to $\nu_n \simeq -1$. This change in number is
consistent with our results, but it also requires an increase in the
characteristic velocity dispersion of 20\%  (i.e., $\nu_\sigma=0.25$).  
For the numbers to decline, the mean mass as indicated by the velocity
dispersion has to increase.  Eliminating this degeneracy requires
samples of lenses large enough to cleanly measure the evolution of the
average image separation with redshift (see \S\ref{sec:sepv}).

Our results on the evolution in the velocity function can also be
compared with recent evolution measurements in the stellar mass function. 
For instance, from the examination of galaxy populations at $z\sim
1-2$ it has been shown that the stellar mass function of galaxies
indeed evolves from $z=0$ to $1$, though the mass dependence on the
form of redshift evolution has yet to be fully clarified
\citep[e.g.,][]{ilbert10,matsuoka10,brammer11}. These studies showed
that the 
number density of galaxies for a given stellar mass range can evolve
by a factor of $\sim 2$ from $z=0$ to $1$, which is in fact
compatible with our results given the large errors on the evolution
parameters for our lensing analysis. While the direct evolution
measurement of the velocity function is very challenging,
\citet{bezanson11} has recently measured the evolution of the velocity
function from $z=0$ to $1.5$ by taking advantage of a scaling
relation between velocity dispersion, stellar mass, and galaxy
structural properties, and found weak redshift evolution up to $z\sim
1$, which is consistent with our result. 

\subsection{Relation between Velocity Dispersion and Image Separation}
\label{sec:sepv}

Another possible source of uncertainties inherent to our analysis 
is the relation between velocity dispersions and image separations.
While this uncertainty is partly taken into account in our analysis
via the systematic error from the dynamical normalization, the true
uncertainties can potentially be larger given complexities such as 
velocity anisotropies and the detailed luminosity profiles of 
galaxies. However, with a large number of lenses we can in principle 
determine the relation from the data, because the ensemble average of
image separations is given by $\langle \theta \rangle \propto
\sigma_*^2$ with a proportionality factor that is almost independent
of cosmological parameters \citep[see][]{kochanek93,kochanek96}.  

\begin{figure}
\epsscale{.55}
\plotone{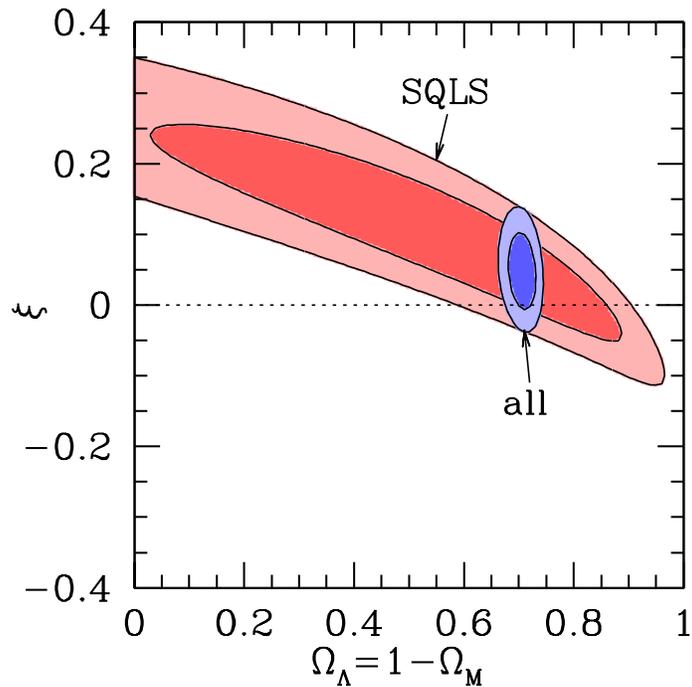}
\caption{Constraints on the cosmological constant $\Omega_\Lambda$ and the
  parameter $\xi$ in Equation~(\ref{eq:sepv}) that parametrizes the
  relation between velocity dispersions and image separations, for
  flat models with a cosmological constant. 
  Outer and inner contours show constraints from the SQLS alone and
  from the combination of all three probes (SQLS+BAO+WMAP),
  respectively.  The horizontal dotted line
  indicates the fiducial value $\xi=0$ assumed in the paper.
  \label{fig:cont_sepv}} 
\end{figure}

To explore the possibility of calibrating the relation between
velocity dispersions and image separations from the lensing data, 
we consider a simple parametric model in which the Einstein radius 
given by Equation~(\ref{eq:tein}) is modified to be
\begin{equation}
\theta_{\rm E}=(1+\xi)^24\pi\left(\frac{\sigma_v}{c}\right)^2
\frac{D_{\rm ls}}{D_{\rm os}},
\label{eq:sepv}
\end{equation}
where the parameter $\xi$ parametrizes the relation such that $\xi=0$
if the velocity dispersion exactly matches the one used for our SIE
models. We consider the flat models with a cosmological constant, and
obtain simultaneous constraints on $\Omega_\Lambda$ and $\xi$. 
The result, shown in Figure~\ref{fig:cont_sepv}, indicates that
constraints on $\xi$ from the SQLS are degenerate with $\Omega_\Lambda$.
The marginalized constraints on each parameter are
$\xi=0.08^{+0.10}_{-0.09}$ and $\Omega_\Lambda=0.62^{+0.20}_{-0.33}$
(statistical errors only). Thus the cosmological constraints become
much weaker, although models with a significant cosmological
constant are still preferred. If we combine all three cosmological
constraints, the resulting constraint is $\xi=0.05^{+0.03}_{-0.03}$.
The slightly positive value reflects the fact that lens statistics
with $\xi=0$ favor slightly larger $\Omega_\Lambda$ than the best-fit
values from BAO and WMAP, and that the observed mean image separation
appears to be slightly higher than the models predict (see
Figure~\ref{fig:sep_dr7stat}). Models with $\xi=0$ are consistent with
the data to better than $2\sigma$.

\section{Conclusion}\label{sec:summary}

We have conducted a statistical analysis of the final sample of strongly 
lensed quasars from the SQLS (Paper V). A subsample of 19
lenses selected from 50,836 quasars has been used to derive
constraints on various cosmological parameters as well as the redshift
evolution of the velocity function of galaxies. We have derived
cosmological constraints assuming no redshift evolution of the
velocity function. For the flat models with a cosmological constant,
we have found 
$\Omega_\Lambda=0.79^{+0.06}_{-0.07}({\rm stat.})^{+0.06}_{-0.06}({\rm syst.})$
from the SQLS alone, which is in concord with other cosmological
constraints. The model with $\Omega_\Lambda=0$ is rejected at $6\sigma$
(statistical errors only), which represents a significant
detection of dark energy independent of Type~Ia supernovae or other
cosmological probes. The systematic error is comparable to the
$1\sigma$ statistical error, suggesting the importance of careful
studies of the systematics for robust cosmological constraints from
lens statistics. We have also derived constraints on non-flat models
with a cosmological constant and flat dark energy models, by combining
the SQLS results with independent constraints from BAO and WMAP, and
obtained results consistent with other studies
\citep[e.g.,][]{komatsu11}. These constraints are summarized in
Table~\ref{tab:cosmo}.   

We have also derived simultaneous constraints on the cosmological
parameters and redshift evolution of the velocity function. 
We parametrize redshift evolution by two parameters $\nu_n$
(Equation~(\ref{eq:evo_nun})) and $\nu_\sigma$ (Equation~(\ref{eq:evo_nus})). 
The SQLS data still prefer a dark energy dominated universe even 
after marginalizing over the evolution parameters, with
$\Omega_\Lambda=0.71^{+0.20}_{-0.25} ({\rm stat.}){}^{+0.12}_{-0.10}({\rm syst.})$
for the flat models with a cosmological constant. We have found no
significant evidence for redshift evolution in the velocity function,
for example $\nu_n=1.06^{+1.36}_{-1.39}({\rm stat.}){}^{+0.33}_{-0.64}({\rm syst.})$ and
$\nu_\sigma=-0.05^{+0.19}_{-0.16}({\rm stat.}){}^{+0.03}_{-0.03}({\rm
  syst.})$, when the SQLS results are combined with BAO and WMAP in the
flat models with a cosmological constant. We summarize the simultaneous
constraints in Table~\ref{tab:evo}. The results remain consistent with
no redshift evolution even if we consider evolution in the
shape of the velocity function. A cautionary note is that because of
the relatively low-redshifts of our source quasars, the SQLS
sample probes galaxy evolution only at $z\la1$.  It is of great
importance to extend the lens statistics 
like the SQLS to higher quasar redshifts in order to study the number
evolution of massive galaxies further, as well as for better
constraints on dark energy. Future wide-field surveys such as
Pan-STARRS and Large Synoptic Survey Telescope will discover thousands
of lensed quasars efficiently by taking advantage of time-domain
information \citep{oguri10b}, which should be helpful for advancing such
applications.

\acknowledgments

This work was supported in part by the FIRST program "Subaru
Measurements of Images and Redshifts (SuMIRe)", World Premier
International Research Center Initiative (WPI Initiative), MEXT,
Japan, and Grant-in-Aid for Scientific Research from the JSPS
(23740161).
This work is supported in part by JSPS Core-to-Core Program
``International Research Network for Dark Energy''.
N.~I. acknowledges support from MEXT KAKENHI 21740151.
M.~A.~S. acknowledges the support of NSF grant AST-0707266.
C.~S.~K. is supported by NSF grant AST-1009756.
C.~E.~R. acknowledges the support of the JSPS Research Fellowship. 
The Institute for Gravitation and the Cosmos is supported by
the Eberly College of Science and the Office of the Senior Vice
President for Research at the Pennsylvania State University.

Funding for the SDSS and SDSS-II has been provided by the Alfred
P. Sloan Foundation, the Participating Institutions, the National
Science Foundation, the U.S. Department of Energy, the National
Aeronautics and Space Administration, the Japanese Monbukagakusho, the
Max Planck Society, and the Higher Education  Funding Council for
England. The SDSS Web Site is http://www.sdss.org/. 

The SDSS is managed by the Astrophysical Research Consortium for the
Participating Institutions. The Participating Institutions are the
American Museum of Natural History, Astrophysical Institute Potsdam,
University of Basel, Cambridge University, Case Western Reserve
University, University of Chicago, Drexel University, Fermilab, the
Institute for Advanced Study, the Japan Participation Group, Johns
Hopkins University, the Joint Institute for Nuclear Astrophysics, the
Kavli Institute for Particle Astrophysics and Cosmology, the Korean
Scientist Group, the Chinese Academy of Sciences (LAMOST), Los Alamos
National Laboratory, the Max-Planck-Institute for Astronomy (MPIA),
the Max-Planck-Institute for Astrophysics (MPA), New Mexico State
University, Ohio State University, University of Pittsburgh,
University of Portsmouth, Princeton University, the United States
Naval Observatory, and the University of Washington.

\clearpage

\begin{deluxetable}{lccccrccl}
\tabletypesize{\footnotesize}
\tablecaption{DR7 Statistical Sample from Paper V\label{tab:sample}}
\tablewidth{0pt}
\tablehead{ 
\colhead{Object} & 
\colhead{$N_{\rm img}$} & 
\colhead{$z_s$\tablenotemark{a}} &
\colhead{$z_l$\tablenotemark{b}} &
\colhead{$i_{\rm PSF}$\tablenotemark{c}} &
\colhead{$\theta_{\rm max}$\tablenotemark{d}} & 
\colhead{$I_{\rm QSO}$ \tablenotemark{e}} & 
\colhead{$I_{\rm gal}$ \tablenotemark{f}} & 
\colhead{Ref.} 
}
\startdata
SDSS~J0246$-$0825 & 2 & 1.686 & 0.723       & 17.76 &  1.09 & 17.80 & 20.78 & 1,2,3 \\
SDSS~J0746+4403   & 2 & 1.998 & 0.513       & 18.71 &  1.08 & 18.71 & 19.62 & 4,5 \\
SDSS~J0806+2006   & 2 & 1.538 & 0.573       & 18.89 &  1.49 & 18.43 & 20.16 & 2,6,7 \\ 
SBS~0909+523      & 2 & 1.378 & 0.830       & 16.17 &  1.11 & 15.94 & 18.81 & 3,8,9 \\ 
SDSS~J0924+0219   & 4 & 1.523 & 0.394       & 18.12 &  1.81 & 18.40 & 19.36 & 3,9,10,11 \\ 
FBQ0951+2635      & 2 & 1.246 & 0.260       & 17.24 &  1.10 & 16.54 & 19.66 & 2,3,12\\ 
SDSS~J1001+5027   & 2 & 1.841 & 0.415       & 17.32 &  2.86 & 17.32 & 19.63 & 13,14 \\ 
SDSS~J1021+4913   & 2 & 1.720 & 0.451       & 18.99 &  1.14 & 18.85 & 19.82 & 14,15 \\ 
SDSS~J1055+4628   & 2 & 1.249 & 0.388       & 18.76 &  1.15 & 18.86 & 19.73 & 5,14 \\ 
PG1115+080        & 4 & 1.735 & 0.311       & 15.97 &  2.43 & 16.40 & 18.91 & 3,16,17\\ 
SDSS~J1206+4332   & 2 & 1.789 & 0.748       & 18.46 &  2.90 & 18.05 & 19.51 & 13  \\ 
SDSS~J1216+3529   & 2 & 2.013 & $0.55\pm0.05$& 19.08&  1.49 & 18.30 & 20.31 & 18  \\ 
SDSS~J1226$-$0006 & 2 & 1.126 & 0.517       & 18.23 &  1.26 & 18.67 & 19.71 & 3,19,20 \\ 
SDSS~J1313+5151   & 2 & 1.877 & 0.194       & 17.70 &  1.24 & 17.10 & 17.49 & 21 \\ 
SDSS~J1335+0118   & 2 & 1.571 & 0.440       & 17.54 &  1.63 & 17.17 & 19.40 & 3,20,22 \\ 
SDSS~J1353+1138   & 2 & 1.624 & $0.25\pm0.05$& 16.47&  1.41 & 16.36 & 17.80 & 6  \\ 
SDSS~J1405+0959   & 2 & 1.810 & 0.66        & 19.05 &  1.98 & 18.70 & 19.70 & 23,24\\  
SDSS~J1455+1447   & 2 & 1.424 & $0.42\pm0.1$& 18.22 &  1.73 & 18.21 & 18.51 & 5\\ 
SDSS~J1515+1511   & 2 & 2.054 & 0.742       & 18.05 &  1.95 & 17.53 & 20.02 & 23\\ \hline
Q0957+561         & 2 & 1.413 & 0.36        & 16.68 &  6.17 & 16.69 & 17.11 & 3,25,26 \\ 
SDSS~J1004+4112   & 5 & 1.740 & 0.68        & 18.82 & 14.72 & 18.44 & 18.82 & 3,27,28,29,30 \\ 
SDSS~J1251+2935   & 4 & 0.802 & 0.410       & 18.86 &  1.79 & 19.32 & 18.43 & 31 \\  
SDSS~J1320+1644   & 2 & 1.502 & 0.899       & 18.88 &  8.59 & 18.36 & 21.59 & 32 \\  
SDSS~J1330+1810   & 4 & 1.393 & 0.373       & 18.35 &  1.76 & 18.34 & 17.84 & 33 \\ 
SDSS~J1332+0347   & 2 & 1.438 & 0.191       & 17.89 &  1.14 & 18.57 & 18.06 & 34\\  
SDSS~J1524+4409   & 2 & 1.210 & 0.320       & 18.76 &  1.67 & 19.49 & 18.33 & 18\\ 
\enddata
\tablecomments{The subsample of 19 lenses above the horizontal solid
  line is used for the final statistical analysis. See text for details.}    
\tablenotetext{a}{Source (quasar) redshifts from the SDSS data.} 
\tablenotetext{b}{Redshifts of lens galaxies. Those with errors are lens redshifts inferred from the color and magnitude of the lens galaxy.} 
\tablenotetext{c}{The PSF magnitude in the SDSS $i$-band magnitude 
 corrected for Galactic extinction. } 
\tablenotetext{d}{Maximum image separation in arcsec.}
\tablenotetext{e}{Johnson $I$-band (Vega) quasar magnitude without
  correcting for Galactic extinction (see text for details).} 
\tablenotetext{f}{Johnson $I$-band (Vega) magnitude of the lens
  galaxy without correcting for Galactic extinction. }
\tablerefs{(1) \citealt{inada05b}; 
           (2) \citealt{eigenbrod07};
           (3) CASTLES webpage (C.~S.~Kochanek et al., http://cfa-www.harvard.edu/castles/.);
           (4) \citealt{inada07};
           (5) \citealt{kayo10};
           (6) \citealt{inada06};
           (7) \citealt{sluse08};
           (8) \citealt{oscoz97};
           (9) \citealt{lubin00};
           (9) \citealt{inada03a};
           (10) \citealt{ofek07};
           (11) \citealt{eigenbrod06a};
           (12) \citealt{schechter98};
           (13) \citealt{oguri05a};
           (14) Paper IV
           (15) \citealt{pindor06};
           (16) \citealt{weymann80};
           (17) \citealt{kundic97};
           (18) \citealt{oguri08b};
           (19) Paper II
           (20) \citealt{eigenbrod06b};
           (21) \citealt{ofek07};
           (22) \citealt{oguri04a};
           (23) Inada et al., in preparation;
           (24) \citealt{jackson12}.
           (25) \citealt{walsh79};
           (26) \citealt{young80};
           (27) \citealt{inada03b};          
           (28) \citealt{oguri04b};          
           (29) \citealt{inada05a};          
           (30) \citealt{inada08b};          
           (31) \citealt{kayo07};
           (32) C.~E. Rusu et al., in preparation; 
           (33) \citealt{oguri08c};
           (34) \citealt{morokuma07}.
           }
\end{deluxetable}

\clearpage

\begin{deluxetable}{cc}
\tablecaption{Breakdown of Systematic Errors for the Flat
  Models with a Cosmological Constant\label{tab:syst}}
\tablewidth{0pt}
\tablehead{
\colhead{Source} & \colhead{Error on $\Omega_\Lambda$}}
\startdata
$f_{\rm obl}\rightarrow \pm 0.25$        & ${}^{+0.037}_{-0.032}$\smallskip \\ 
$\bar{e}\rightarrow \pm 0.1$                & ${}^{+0.000}_{-0.006}$ \smallskip\\
$z_s$ for $\kappa_{\rm ext}$ and $\gamma_{\rm ext}$ 
$\rightarrow \pm1$ & ${}^{+0.003}_{-0.000}$ \smallskip\\
$\beta_{\rm l}\rightarrow \pm 0.1$          & ${}^{+0.019}_{-0.018}$ \smallskip\\
Different $dn/d\sigma_v$                    & ${}^{+0.000}_{-0.046}$ \smallskip\\
$dn/d\sigma_v$ evolution                    & ${}^{+0.033}_{-0.000}$ \smallskip\\ 
$i_{\rm QSO}-i_{\rm gal}\rightarrow\pm0.25$ & $^{+0.019}_{-0.011}$ \smallskip\\ 
\hline 
Total & ${}^{+0.057}_{-0.060}$ \smallskip\\
\enddata
\tablecomments{The total errors are estimated from the quadrature sum
  of all the systematic errors. }
\end{deluxetable}

\clearpage

\begin{deluxetable}{llccc}
\tablecaption{Constraints on Cosmological Parameters\label{tab:cosmo}}
\tablewidth{0pt}
\tablehead{
\colhead{Model} & \colhead{Data} & \colhead{$\Omega_M$} &
\colhead{$\Omega_{\rm DE}$} & \colhead{$w$}}
\startdata
flat $\Omega_\Lambda$    & SQLS     & $\equiv 1-\Omega_\Lambda$ & $0.79^{+0.06}_{-0.07}{}^{+0.06}_{-0.06}$ & $\equiv -1$\\ \hline
non-flat $\Omega_\Lambda$& SQLS+BAO & $0.28^{+0.03}_{-0.03}{}^{+0.02}_{-0.02}$ & $0.88^{+0.09}_{-0.10}{}^{+0.07}_{-0.09}$ & $\equiv -1$\\ 
non-flat $\Omega_\Lambda$& SQLS+WMAP& $0.20^{+0.08}_{-0.06}{}^{+0.07}_{-0.07}$ & $0.78^{+0.05}_{-0.06}{}^{+0.05}_{-0.05}$ & $\equiv -1$\\ 
non-flat $\Omega_\Lambda$& SQLS+BAO+WMAP& $0.29^{+0.02}_{-0.02}{}^{+0.00}_{-0.00}$ & $0.71^{+0.02}_{-0.02}{}^{+0.00}_{-0.00}$ & $\equiv -1$\\ \hline
flat $\Omega_{\rm DE}$   & SQLS+BAO & $0.25^{+0.03}_{-0.03}{}^{+0.03}_{-0.02}$ & $\equiv 1-\Omega_M$  &  $-1.44^{+0.22}_{-0.25}{}^{+0.17}_{-0.18}$\\
flat $\Omega_{\rm DE}$   & SQLS+WMAP& $0.23^{+0.04}_{-0.03}{}^{+0.03}_{-0.03}$ & $\equiv 1-\Omega_M$  &  $-1.19^{+0.17}_{-0.17}{}^{+0.14}_{-0.15}$\\
flat $\Omega_{\rm DE}$   & SQLS+BAO+WMAP& $0.28^{+0.02}_{-0.02}{}^{+0.01}_{-0.01}$ & $\equiv 1-\Omega_M$ & $-1.11^{+0.14}_{-0.17}{}^{+0.08}_{-0.10}$\\
\enddata
\tablecomments{The parameter values are followed by their statistical
  and systematic uncertainties, respectively. See the last paragraph of
  \S~\ref{sec:intro} for the definitions of cosmological parameters.}    
\end{deluxetable}

\clearpage

\begin{deluxetable}{llcccrc}
\tablecaption{Constraints on Cosmological Parameters and the Redshift
  Evolution of the Velocity Function\label{tab:evo}}
\rotate
\tablewidth{0pt}
\tablehead{
\colhead{Model} & \colhead{Data} & \colhead{$\Omega_M$} &
\colhead{$\Omega_{\rm DE}$} & \colhead{$w$} & \colhead{$\nu_n$} & \colhead{$\nu_\sigma$}}
\startdata
flat $\Omega_\Lambda$    & SQLS         & $\equiv 1-\Omega_\Lambda$ & $0.71^{+0.20}_{-0.25}{}^{+0.12}_{-0.10}$ & $\equiv -1$ & $1.07^{+1.61}_{-1.64}{}^{+0.27}_{-0.82}$ & $-0.05^{+0.20}_{-0.20}{}^{+0.04}_{-0.05}$\\ 
flat $\Omega_\Lambda$    & SQLS+BAO     & $\equiv 1-\Omega_\Lambda$ & $0.69^{+0.03}_{-0.04}{}^{+0.00}_{-0.00}$ & $\equiv -1$ & $1.15^{+1.36}_{-1.43}{}^{+0.33}_{-0.72}$ & $-0.05^{+0.19}_{-0.16}{}^{+0.03}_{-0.03}$\\ 
flat $\Omega_\Lambda$    & SQLS+WMAP    & $\equiv 1-\Omega_\Lambda$ & $0.73^{+0.02}_{-0.03}{}^{+0.00}_{-0.00}$ & $\equiv -1$ & $1.00^{+1.35}_{-1.41}{}^{+0.30}_{-0.63}$ & $-0.06^{+0.19}_{-0.16}{}^{+0.04}_{-0.03}$\\ 
flat $\Omega_\Lambda$    & SQLS+BAO+WMAP& $\equiv 1-\Omega_\Lambda$ & $0.71^{+0.02}_{-0.02}{}^{+0.00}_{-0.00}$ & $\equiv -1$ & $1.06^{+1.36}_{-1.39}{}^{+0.33}_{-0.64}$ & $-0.05^{+0.19}_{-0.16}{}^{+0.03}_{-0.03}$\\ \hline
non-flat $\Omega_\Lambda$& SQLS+BAO     & $0.25^{+0.06}_{-0.05}{}^{+0.01}_{-0.01}$ & $1.02^{+0.20}_{-0.28}{}^{+0.06}_{-0.03}$ & $\equiv -1$ & $0.15^{+1.61}_{-1.39}{}^{+0.20}_{-0.34}$ & $-0.14^{+0.20}_{-0.23}{}^{+0.08}_{-0.02}$ \\ 
non-flat $\Omega_\Lambda$& SQLS+WMAP    & $0.32^{+0.11}_{-0.09}{}^{+0.04}_{-0.13}$ & $0.69^{+0.07}_{-0.08}{}^{+0.10}_{-0.03}$ & $\equiv -1$ & $1.29^{+1.24}_{-1.62}{}^{+0.38}_{-0.66}$ & $-0.06^{+0.21}_{-0.16}{}^{+0.02}_{-0.06}$ \\ 
non-flat $\Omega_\Lambda$& SQLS+BAO+WMAP& $0.30^{+0.02}_{-0.02}{}^{+0.00}_{-0.00}$ & $0.71^{+0.02}_{-0.01}{}^{+0.00}_{-0.00}$ & $\equiv -1$ & $1.06^{+1.22}_{-1.24}{}^{+0.32}_{-0.65}$ & $-0.05^{+0.15}_{-0.16}{}^{+0.04}_{-0.03}$ \\ \hline
flat $\Omega_{\rm DE}$   & SQLS+BAO     & $0.21^{+0.07}_{-0.05}{}^{+0.01}_{-0.01}$ & $\equiv 1-\Omega_M$ & $-1.70^{+0.45}_{-0.40}{}^{+0.05}_{-0.10}$ & $-0.48^{+1.71}_{-1.56}{}^{+0.33}_{-0.29}$ & $-0.04^{+0.19}_{-0.19}{}^{+0.04}_{-0.03}$ \\
flat $\Omega_{\rm DE}$   & SQLS+WMAP    & $0.17^{+0.15}_{-0.06}{}^{+0.06}_{-0.05}$ & $\equiv 1-\Omega_M$ & $-1.51^{+0.69}_{-0.60}{}^{+0.34}_{-0.32}$ & $-0.51^{+2.38}_{-2.18}{}^{+0.88}_{-0.80}$ & $-0.07^{+0.17}_{-0.19}{}^{+0.04}_{-0.03}$ \\
flat $\Omega_{\rm DE}$   & SQLS+BAO+WMAP& $0.29^{+0.02}_{-0.03}{}^{+0.00}_{-0.01}$ & $\equiv 1-\Omega_M$ & $-1.01^{+0.14}_{-0.18}{}^{+0.03}_{-0.03}$ & $ 1.06^{+1.18}_{-1.39}{}^{+0.28}_{-0.66}$ & $-0.05^{+0.16}_{-0.14}{}^{+0.03}_{-0.02}$ \\
\enddata
\tablecomments{The parameter values are followed by their statistical
  and systematic uncertainties, respectively. See the last paragraph of
  \S~\ref{sec:intro} for the definition of cosmological parameters.}    
\end{deluxetable}

\clearpage

\clearpage

\clearpage

\clearpage

\clearpage

\clearpage

\clearpage

\clearpage

\clearpage

\clearpage

\clearpage

\clearpage

\end{document}